\documentclass{article}

\usepackage{arxiv}

\usepackage[utf8]{inputenc} 
\usepackage[T1]{fontenc}    
\usepackage{hyperref}       
\usepackage{url}            
\usepackage{booktabs}       
\usepackage{amsfonts}       
\usepackage{nicefrac}       
\usepackage{microtype}      
\usepackage{lipsum}

\usepackage{graphicx}
\usepackage{amssymb}
\usepackage{amsmath}
\usepackage{multirow}

\usepackage[ruled,vlined]{algorithm2e}
\usepackage{algorithmic}

\title{Constrained plasticity reserve as a natural way to control frequency and weights in spiking neural networks}

\author{
  Oleg~Nikitin\\
  Computing Center \\
  Far Eastern Branch\\ 
  Russian Academy of Sciences\\
  680000, Khabarovsk, Russia\\
  \texttt{olegioner@gmail.com} \\
   \And
 Olga~Lukyanova\thanks{Corresponding author} \\
  Computing Center \\
  Far Eastern Branch\\ 
  Russian Academy of Sciences\\
  680000, Khabarovsk, Russia\\
  \texttt{ollukyan@gmail.com} \\
  \And
 Alex~Kunin \\
  Computing Center \\
  Far Eastern Branch\\ 
  Russian Academy of Sciences\\
  680000, Khabarovsk, Russia\\
  \texttt{alexkunin88@gmail.com} \\
}

\begin{document}
\maketitle

\begin{abstract}
Biological neurons have adaptive nature and perform complex computations involving the filtering of redundant information. However, most common neural cell models, including biologically plausible, such as Hodgkin-Huxley or Izhikevich, do not possess predictive dynamics on a single-cell level. Moreover, the modern rules of synaptic plasticity or interconnections weights adaptation also do not provide grounding for the ability of neurons to adapt to the ever-changing input signal intensity. While natural neuron synaptic growth is precisely controlled and restricted by protein supply and recycling, weight correction rules such as widely used STDP are efficiently unlimited in change rate and scale. The present article introduces new mechanics of interconnection between neuron firing rate homeostasis and weight change through STDP growth bounded by abstract protein reserve, controlled by the intracellular optimization algorithm. We show how these cellular dynamics help neurons filter out the intense noise signals to help neurons keep a stable firing rate. We also examine that such filtering does not affect the ability of neurons to recognize the correlated inputs in unsupervised mode. Such an approach might be used in the machine learning domain to improve the robustness of AI systems.
\end{abstract}





\section{Motivation}\label{sec:1}

Biological neural networks are different from digital analogs in many ways. Brain neurons are slow, noisy, and constrained by energy and spatial dimensions. However, they still outperform modern AI systems in many ways. Self-learning, robustness, and few-shot learning are among the edges of natural brains \cite{29}. Most of the present neural networks practically used in machine learning are based on the point neuron model. Neurons are just summation devices in these neural nets. In the present article, we would like to consider some intracellular features of neurons that regulate their activity and lead to better learning. 

\begin{figure}
\centering
\includegraphics[width=0.5\textwidth]{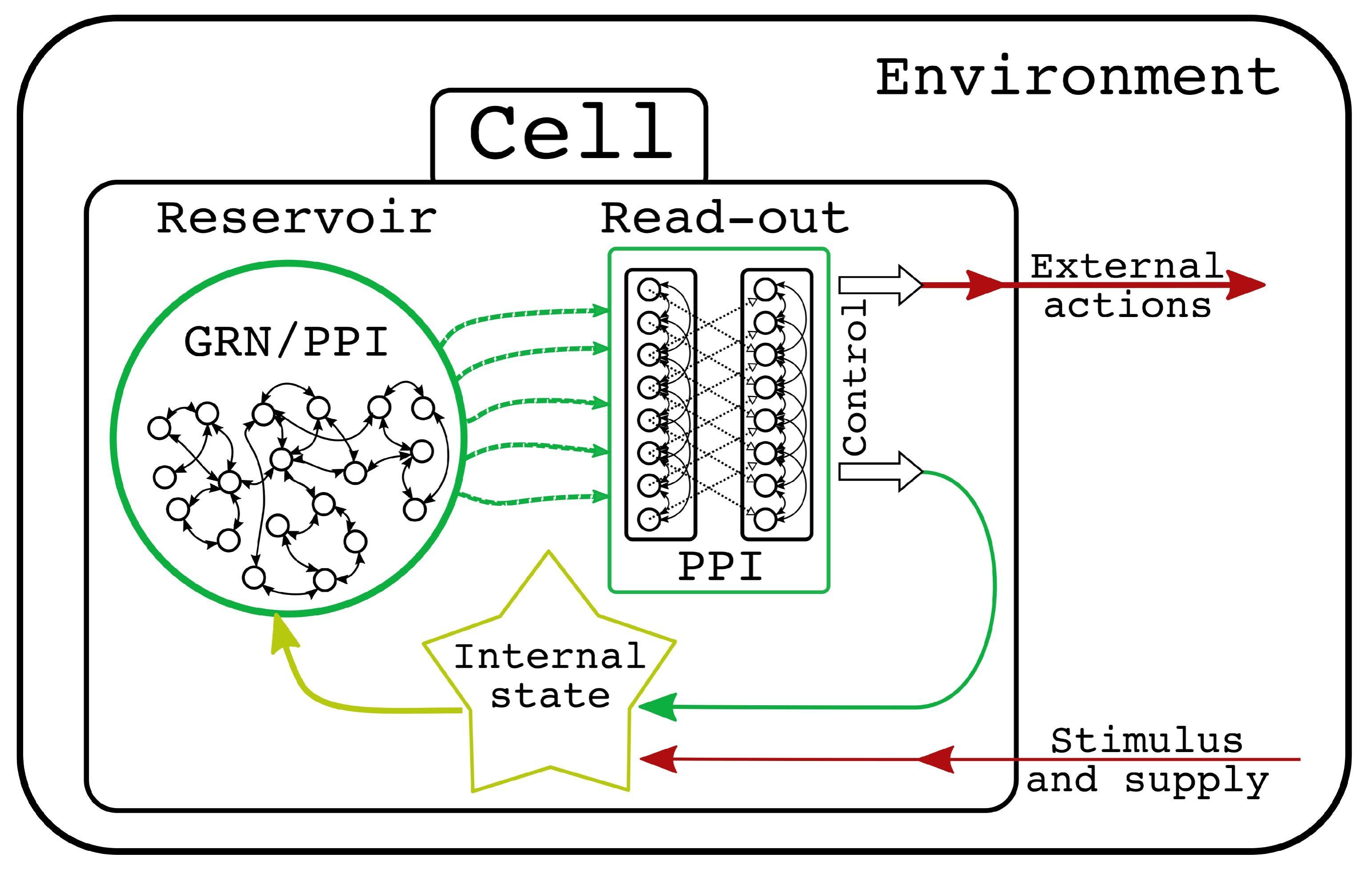}
\caption{The conceptual structure of a biological cell as a reservoir computing device.}
\label{fig:1}
\end{figure}

Biological cells possess a molecular and genetic substrate for complex adaptation and regulation and perform different actions and decisions to adapt to the dynamic environment. Such adaptive structures are present as Genetic regulatory networks (GRN) and Protein-protein interaction (PPI) networks \cite{1} that can be viewed as reservoir computer or approximation device, based on the interaction between the recurrent layer and readout layer (Figure~\ref{fig:1}). GRN and PPI networks of Escherichia coli were modeled in several studies \cite{2,4,5}. In the present article, we abstract GRNs and PPIs into the computational mechanisms that regulate the neural activity toward the desired firing rate. Neurons should keep their firing rate on a more or less constant level and compensate for overexcitation. The firing rate homeostasis can be achieved by predictive up or down-regulation of cellular activity. This process involves control of synthesis and distribution of different proteins, allowing cells to let in the more or less extracellular excitatory or inhibitory stimulus.

We want to proceed by the way proposed in \cite{29} or build less 'artificial' neural networks with bio-inspiration. Therefore, we use the spiking neuron model not only because it is biologically more plausible but also because it has some meaningful advantages. For example, spiking neural nets are efficient in energy consumption and memory allocation, better suited for sequential data processing than regular neural nets \cite{Deng}. Spiking neurons might be a good model for implementing predictive learning approaches \cite{Masumori}. 

Also, STDP learning rules have a significant advantage in application to neuromorphic hardware. These computational models may be implemented in analogous electronics, which will lead to fast and energy-efficient computations. \cite{Yao} reviewed different implementations of STDP in hardware. In the present paper, we would like to study the firing rate-related regulation of STDP weight growth. \cite{Wang} showed control of the intensity of STDP in InGaZnO memristors by varying the duration time of potentiation. This feature creates an opportunity to implement rate-based restrictions for STDP growth by linking the firing rate of postsynaptic neuron to the time of applying voltage current to memristive plasticity.

With regular artificial neural networks trained by backpropagation of error, it is hard to implement recurrent architectures \cite{Bengio}. Most importantly, the 'traditional' artificial neural network (ANN) may act only reactive towards input signals. Despite the advantages of spiking nets, they are not well suited for backpropagation algorithms due to the non-differentiable nature of the network signals. At the same time, Hebbian learning \cite{Hebb} approaches such as spike-timing-dependent plasticity (STDP) provide suitable rules for local unsupervised learning. STDP type of learning is well studied, and it is good at finding coincidences in the data \cite{Gerstner}.

\begin{figure}
\centering
\includegraphics[width=0.5\textwidth]{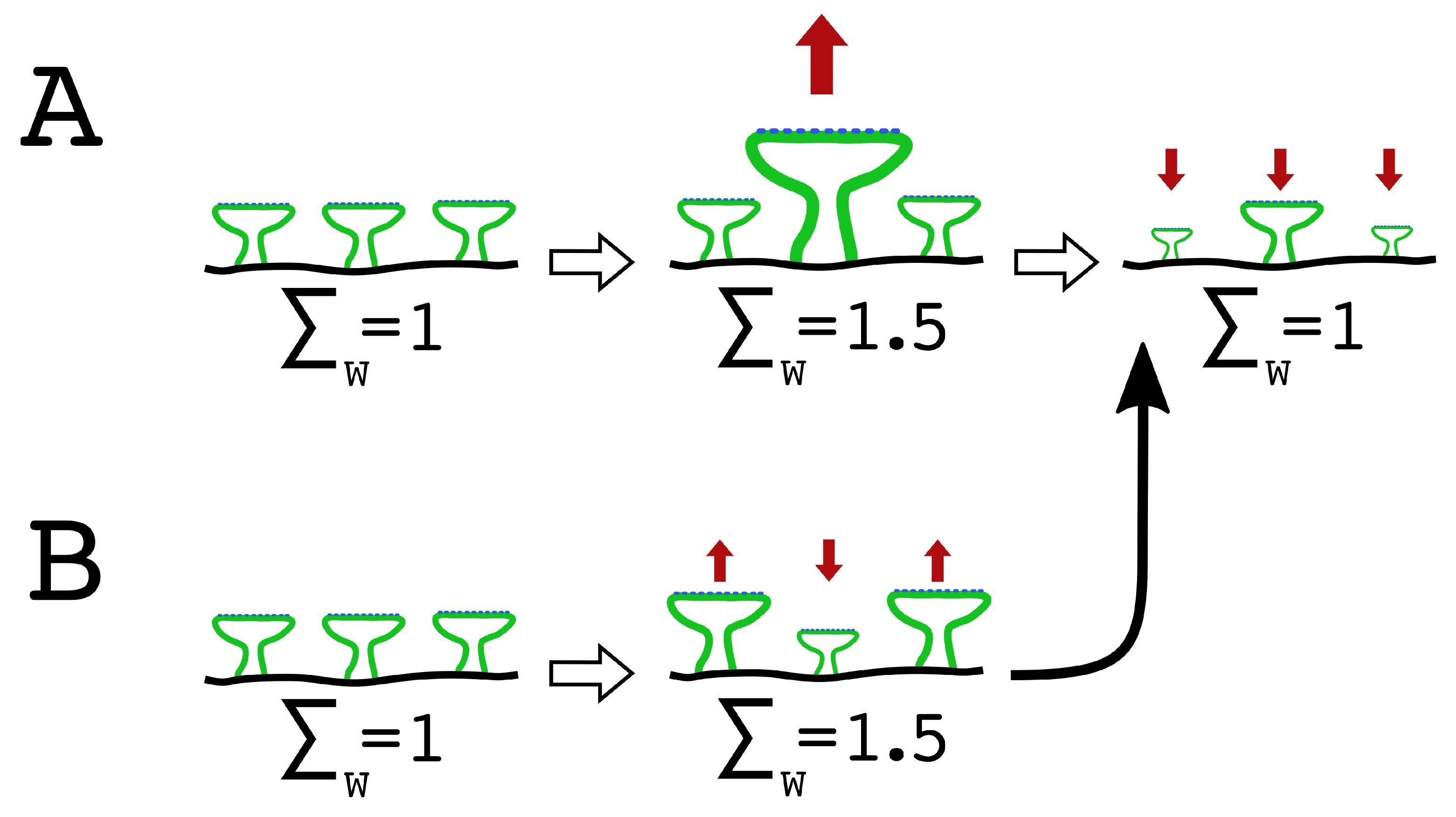}
\caption{Synaptic connections dynamics during processes of (A) homeostatic synaptic scaling, (B) heterosynaptic scaling. On (A), the growth of one synapse is then compensated by decreasing all synapses sizes to maintain the stable 'sum' of all connections weights in a particular region of a dendrite. The timescale of the process is quite long and may last for hours. On (B), the extensive growth of most synapses in one area leads to instant shrinkage of the less stimulated synapses. This process is not keeping the 'sum' of synapses constant and may be continued by (A).}
\label{fig:2}
\end{figure}

\cite{38} showed that the STDP-only rule with no constraints would lead to runaway activity. \cite{35} found that homeostatic scaling constraints biological synapses (Figure~\ref{fig:2}A) when all the weights are dynamically rescaled to keep the firing rate of a postsynaptic neuron on a constant level. Heterosynaptic scaling accompanies homeostatic scaling \cite{37} (Figure~\ref{fig:2}B) when growth (or long-term potentiation, LTP) of one synapse leads to a decrease (or long-term depression, LTD) of other synapses. Thus, it introduces the synaptic competition and leads to the amplification of the most important inputs.

Homeostatic and heterosynaptic scalings are both essential for the stability and reliable work of natural neural systems. The work of \cite{DiPaolo} was one of the first works considering fast-spiking penalization for the rate-based plasticity and showed the practical applicability of rate-controlled plasticity for the adaptive control task. \cite{VanRossum} first simulated the computational model of rate-based restrictions of plasticity for STDP known as Homeostatic synaptic scaling (HSS). \cite{Zenke} demonstrated that HSS alone would lead to unstable network behavior and that STDP weight growth should be constrained by frequency-dependent regulation. \cite{38} and \cite{36} proposed a computational model of weight restraining rule leading to the heterosynaptic regulation. It was implemented basically as HSS that works only for most extreme (small or big) weights and only when the firing frequency of a neuron is too high, which prevents neurons from runaway activity and leads to the potentiation of some synapses against another. However, it is unclear from the model if it preserves the most correlated and essential connections.

We propose including some features of cellular self-org-\allowbreak anization into the STDP weight update rule to improve the spiking neural network (SNN) ability to process the dynamic input signals. We want to investigate into more 'physically' inspired approach leading to synaptic competition. Natural synaptic weight growth is tightly constrained by construction 'materials' (proteins) availability and precisely controlled regarding maintaining stable cellular activity \cite{33, 34}. Below, we propose the model, which naturally regulates the weight growth in all the compartmentalized artificial neural cell parts by the control of reserve of abstract 'materials' for potential synaptic weight growth. This type of restriction is nonlinear and may naturally regularize the growth of weights in the STDP rule.

\section{The neuron and plasticity model overview}\label{sec:2}

Basic implementations of STDP incorporate temporal coincidence detection and well reproduce the Hebbian plasticity principle. However, it tends to unbounded growth of some weights. Even the introduction of stabilizing term leads to the development of the bimodal distribution, with some weights growing to maximum values and others bounding around zero. In the current paper, we propose the novel model of control of STDP weights, which will improve its ability to learn and will introduce the control of firing rate in the networks. This work is a continuation of previous research on activity homeostasis in neural networks \cite{Procedia18}, where authors proposed a model of a neural cell, which allows switching behavioral programs, maintaining a stable cell functioning, and responding to changes in the dynamics of input signals. A neural network with homeostatic dependent input-switching activity was shown in \cite{LNCS18}. The features of the model allowed it to change the behavior activity adaptively.

Based on the research discussed above, we propose the synaptic plasticity regulation model. We introduce the concept of 'plasticity potential reserve' as generalized abstract 'materials' for the potential growth of synaptic weight in our model. We use it to regulate the processes of weight growth. We suppose that to grow weights neuron needs to spend some material substances. Neuron controls the value of the plasticity potential reserve available for weight growth, affecting the synaptic plasticity and stabilizing the neuron activity for the desired firing rate dynamically.

In the present paper, we use a model of one spiking neuron with dendrites and an activation model based on \cite{Izhikevich}. 

We will consider neuron $n$ with $m$ synaptic inputs distributed across $D$ dendritic compartments such that each dendrite has $S$ synapses. The number of synapses is the same for each dendrite. As we consider only one neuron we can discard $n$ term and operate only with $m$ weights $w_m$ across $D$ dendrites. Further, we uniquely map $w_{nm}$ to $w_{ds}$. So, we define $w_{ds}$ as a current synaptic weight of the synapse $s$ ($s \in S$) on the dendrite $d$ ($d \in D$) in time $t$.

For the neuron's activation function, we choose the Izhikevich model because of its biological plausibility, low computational complexity, and inherent adaptivity. It should be noted that other activation functions such as Leaky integrate-and-fire (LIF) or Spike response model (SRM) might be considered for future research \cite{Kistler}. All weighted inputs are integrated first at dendrites $D$, and the signals from dendrites are summed on the somatic body of the neuron denoted as $I$ and projected to the Izhikevich model (Equation~\ref{eq:a1} in Appendix~\ref{subsec:a1}).

Here, the input current $I$ is the sum of all the received inputs $I_{ds}$ to each synapse $s$ on the dendrite $d$ multiplied by all synaptic weights $w_{ds}$, respectively. It is averaged over synapses and then over dendrites and scaled by the input scaling coefficient:

\begin{equation}\label{eq:1}
I = \frac{1}{D}\cdot \sum\limits_{d}\left(\frac{2\cdot \sum\limits_{s}{(I_{ds} \cdot w_{ds})}}{S\cdot (w_{max} - w_{min})}\right)\cdot k^{izh},
\end{equation}
where $w_{max}$ and $w_{min}$ are the maximum and the minimum possible weights from the Equation~\ref{eq:a14} in Appendix~\ref{subsec:a2}; $k^{izh}$ is the input scaling coefficient, described in Appendix~\ref{subsec:a1}, Equations~\ref{eq:a3}--\ref{eq:a12}.

According to Izhikevich equations, the input values $I$ are mapped into voltage $v$ of a neuron through nonlinear transformations. Neuron produces spike output $O$ when $v$ exceeds threshold value $v_{peak}$.

The model should function with the same dynamics with no respect to changes in the number of dendrites and synapses on each dendrite. So, we need to compute activations in the relative form to achieve the scalability of the functioning of our model neuron. We propose to compute the input current by averaging across synapses and dendrites so that the changes in dimensions of the model will not affect the activation function dynamics. Otherwise, 'bigger' neurons will lead to faster saturation and spiking of the Izhikevich model, which would lead to uncontrolled behavior. We also scale the input activity by multiplying on scaling coefficient $k^{izh}$. This scaling coefficient is derived from general Izhikevich equations and will keep the inputs in the desired physiologically adequate range despite the potential fluctuations in the input intensity. The derivation procedure of $k^{izh}$ is provided in Appendix~\ref{subsec:a1}.

In our model of the neuron, synapses are plastic, and their basic dynamic $\Delta w^{stdp}_{ds}$ (Appendix\ref{subsec:a2}) is governed by the STDP rule introduced in \cite{STDP}. The present paper is devoted to studying the approaches to regularize the weight growth in STDP-based plasticity. The essence of the approaches is represented by controlling STDP weight growth. The following sections will cover the approaches to the synaptic constraints leading to firing rate homeostasis. All the restrictions terms described in the following sections are applied to the $\Delta w^{stdp}_{ds}$ value. 

However, other types of plasticity regulation, different from firing rate-dependent, exist in neural systems. Here we would like to discuss the heterosynaptic plasticity and an approach to simulate it. \cite{GonzalezIslas} and \cite{Fong} showed that synaptic stabilization is performed in local postsynaptic compartments independent from the firing frequency. This process is not leading to any frequency adaptation. It cannot be explained solely by homeostatic plasticity models, like one in \cite{VanRossum}. It seems to be determined only by local processes in cellular compartments. In our model, we have separate dendrites; thus, we may try to reproduce the phenomenon from \cite{Fong}. We will call our model of dendritic weight stabilization a Spike-independent synaptic scaling (SISS). The SISS can be performed on separate dendritic compartments, which means that the sum of all weights on one particular dendrite is aimed to be constant. The goal weight $w^{ideal}_{ds}$ is set for each particular synapse, and then the scaling coefficient for all synapses on the dendrite is calculated according to Equation~\ref{eq:2}:

\begin{equation}\label{eq:2}
k^{siss}_{d} = \frac{\sum\limits_{s}{w_{ds}}\cdot (\tau_{siss} - 1)+\sum\limits_{s}w^{ideal}_{ds}}{\sum\limits_{s}{w_{ds}}\cdot \tau_{siss}},
\end{equation}
where $w^{ideal}_{ds}$ is the set of weights values to the sum of which the sum of weights on the dendrite $d$ seeks to approach; and $\tau_{siss}$ is the time constant, determining the number of steps the weight should take to reach the $w^{ideal}_{ds}$.

In the present paper, we set the value of $\tau_{siss}$ to 10 and $w^{ideal}_{ds}$ to 0.5 for all the synapses. Note that $w^{ideal}_{ds}$ could differ for each synapse and may be updated during the simulation as an arbitrary matrix for all synapses of the dendrite. One possible scenario may be setting the $w^{ideal}_{ds}$ as a $w_{ds}$ on the particular time step as a result of some special event, which may lead to more flexible usage of SISS, leading to some different points during the simulation. However, below, we will focus only on the constant type of $w^{ideal}_{ds}$ for the sake of simplicity.

In the following sections, we will describe several approaches to weight growth control. First, we will start from a simple linear form, known as synaptic scaling (or homeostatic scaling) \cite{35}, which below we will refer to as HSS. Then, we will describe the implementation of heterosynaptic plasticity based on the physical supply of synaptic growth materials. We refer to these materials as plasticity potential. Next, we will introduce the models of synthesis of plasticity potential in soma and distribution of it into dendrites and synapses onward. Then we will propose three approaches to the control of the plasticity potential secretion in the soma of a neuron, one retrospective and two forward-looking.

\section{Approaches to homeostatic synaptic plasticity}\label{sec:3}

The neurons of the brain have a quite strict range of normal functioning in terms of firing rate. Each neuron keeps some normal physiological level of spiking activity that it has to obey to avoid cellular death and maintain efficient network cooperation. These optimal firing rates vary for different cell types and locations in the brain. They define the set points of homeostasis that govern cellular activity. In biological neural networks, the mitochondrial activity defines the exact setpoints \cite{Styr}. In our case, the target firing rate $\theta_{target}$ is set by hand, while in further research, the optimization techniques for optimal firing rates of the neurons should be considered. Below, we will choose the target firing rate $\theta_{target}$ on the 'low' and 'high' activity levels. The exact values may reflect the input signal regimes. For example, the input signal with a 'low' average rate of 0.2 may be accompanied by $\theta_{target}$ of 0.2, and it will be considered a 'normal' target rate. In the case of incoming 'high' frequency input with a rate of 0.5, we may keep the 'low' target rate to study how the neuron will perform in the presence of overstimulation. Our model then should exhibit the high-pass filtering features.

Below, we will study different approaches to weights' regularization based on firing rate homeostasis. We will consider a linear approach to modeling homeostatic synaptic plasticity, such as HSS, and a nonlinear one -- the authors' approach to synaptic homeostasis based on the control of the plasticity potential leading to the heterosynaptic competition. Further, they will be studied and compared in terms of efficiency and applicability for firing rate control tasks.

\subsection{HSS}\label{subsec:31}

Here, we implement a simple version of an approach to homeostatic synaptic plasticity described in \cite{VanRossum}. The basic idea is to scale all weights of a neuron in response to firing frequency changes. 

After adding $\Delta w^{stdp}_{ds}$, synaptic weights ${w}_{ds}$ are corrected depending on the deviation of spiking frequency from the target:

\begin{equation}\label{eq:3}
\dot{w}_{ds} = (1 - (\theta_{real} - \theta_{target})) \cdot \left(w_{ds}+\Delta w^{stdp}_{ds} \right),
\end{equation}
where $\theta_{real}$ is the current firing frequency and $\theta_{target}$ is the target firing frequency.

Here, $\theta_{target}$ is a given target constant and $\theta_{real}$ is a simple moving average (SMA) calculated as:

\begin{equation}\label{eq:4}
\theta_{real} = \frac{\sum\limits^{t}_{t-wind}O(t)}{wind}
\end{equation}
where $O(t)$ is the action potential generation (spiking) at the time $t$, which can be '0' or '1', and $wind$ is the given window, showing how many steps back we are looking.

The further the value of the $\theta_{real}$ is from the value of the $\theta_{target}$, the proportionally this decreases $\Delta w^{stdp}_{ds}$ in the directions that can increase the difference between these frequencies. Thus, this rule holds back the distance between 'real' and 'target' firing frequencies. 

The presented HSS approach set a linear restriction on all the input weights. While for efficient learning, it is beneficial to reinforce the most important signals and filter an effective noise. In the next section, we describe the nonlinear dynamic approach to weight outgrowth control.

\subsection{Synaptic homeostasis based on control of plasticity potential}\label{subsec:32}

\begin{figure*}
\centering
\includegraphics[width=0.9\textwidth]{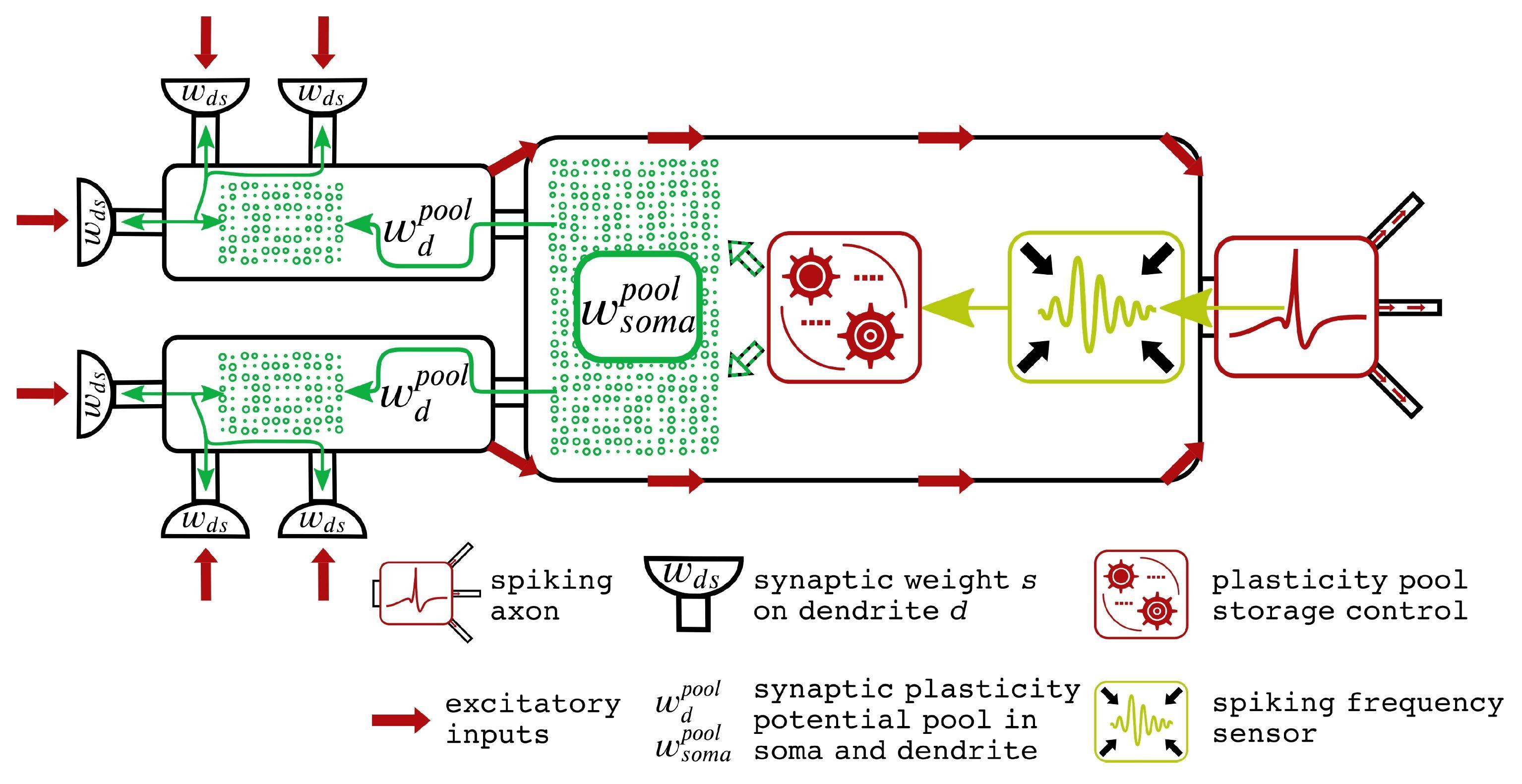}
\caption{Schematic representation of proposed weight control model.}
\label{fig:3}
\end{figure*}

As we described above in the Section~\ref{sec:1} of the paper, biological neurons do control the weight growth and firing frequency by complex intracellular GRN/PPI networks. Below, we introduce an abstract model of such a control approach based on the availability of abstract 'materials' required for weight growth. The present approach allows for natural competition between synapses on dendrites and competition between different dendrites. The model schematically showed in Figure~\ref{fig:3}. The three main parts of the system are following:

\begin{enumerate}
\item the reserve of synaptic weight growth in the soma ($w^{pool}_{soma}$) and in dendrites ($w^{pool}_d$); 
\item the action potential generation model (spiking model) and firing frequency detector; 
\item the update algorithm for the reserve of synaptic weight growth with regard to firing frequency control. 
\end{enumerate} 

The reserve of synaptic weight growth moves unidirectionally from the soma to dendrites. In the dendrites, the growth materials are used during LTP and partly returned to dendrite during LTD. Therefore, the particular implementation of the control algorithm might be different, ranging from simple frequency-dependent control or some black box control system or the differential equations for optimal control. Below, we describe the additions and extensions to STDP that introduce some nonlinearity and weight growth control.

Consider the dynamics of the synaptic weights $w_{ds}$ by the dendrite $d$ and synapse $s$ respectively: 

\begin{equation}\label{eq:5}
\dot w_{ds} = k^{siss}_{d} \cdot \left(w_{ds}+\Delta w^{stdp}_{ds} \right),
\end{equation}
where $\dot w_{ds}$ is the dynamics in the synaptic weight of the synapse $s$ on the dendrite $d$ with respect to time; $k^{siss}_{d}$ is the spike-independent synaptic scaling coefficient on the dendrite $d$; $w_{ds}$ is the current synaptic weight of the synapse $s$ on the dendrite $d$ in time $t$; $\Delta w^{stdp}_{ds}$ is the update of the synaptic weights by spike-timing-dependent plasticity of the synapse $s$ on the dendrite $d$.

In our proposed model of weight growth regulation depending on the available pool of plasticity potential resources, we will restrict the growth of calculated $\Delta w^{stdp}_{ds}$ depending on the availability of growth resources $w^{pool}_{d}$ in the dendrite $d$. Thus, we add the restriction coefficient of the synaptic weight growth $k^{wp}_{d}$:

\begin{equation}\label{eq:6}
\Delta w^{stdp}_{ds} =
\begin{cases} 
\hfill \Delta w^{stdp'}_{ds}, & \text{if } \Delta w^{stdp'}_{ds} < 0, \\
k^{wp}_{d} \cdot \Delta w^{stdp'}_{ds}, & \text{if } \Delta w^{stdp'}_{ds} > 0, \\
\end{cases}
\end{equation}
where $k^{wp}_{d}$ is the restriction coefficient of the growth of synaptic weights depending on the deviation from the optimum frequency.

\begin{equation}\label{eq:7}
\begin{aligned}
k^{wp}_{d} = {} & 1 + \left( \frac{w^{pool}_{d}}{k^{siss}_{d} \cdot \sum\limits_{s}{\Delta w^{stdp+}_{ds}}} - 1 \right) \cdot \\
& \cdot H \left( \sum\limits_{s}{\Delta w^{stdp+}_{ds}} - \frac{w^{pool}_{d}}{k^{siss}_{d}} \right),
\end{aligned}
\end{equation}
where $w^{pool}_{d}$ is the potential reserve of synaptic weight growth on the dendrite $d$; $\sum\limits_{s}{\Delta w^{stdp+}_{ds}}$ is the sum by the synapse $s$ of all positive STDP updates of the synaptic weights, and $\sum\limits_{s}{\Delta w^{stdp+}_{ds}} = \sum\limits_{s}{(k^{wp}_{d} \cdot \Delta w^{stdp}_{ds})}$, where $\Delta w^{stdp}_{ds} > 0$; \\ $H \left( \sum\limits_{s}{\Delta w^{stdp+}_{ds}} - \frac{w^{pool}_{d}}{k^{siss}_{d}}\right)$ is a Heaviside step function $H(x)$, also known as the 'unit step function', and written as

\begin{equation}\label{eq:8}
H(x)=
\begin{cases}
0, & \text{if}\ x<0 \\
1, & \text{if}\ x>0 \\
\end{cases}
\end{equation} 

The coefficient $k^{wp}_{d}$ is in the range from 0 to 1. In the case when the sum of all positive changes in weights ($\sum\limits_{s}{\Delta w^{stdp+}_{ds}}$) on the dendrite is less than or equal to the value of the dendritic plasticity pool ($w^{pool}_{d}$), $k^{wp}_{d}$ will take its maximum value equal to 1. In other words, the amount of $w^{pool}_{d}$ on the dendrite is sufficient to change the weights positively. If it is not enough, then $k^{wp}_{d}$ will be less than 1; that is, a positive change in weights will occur partially.

Thus, we can calculate the dynamics of synaptic weights using the STDP and SISS, which have different effects on the resulting value of synaptic weights (increase or decrease), and determines the potential reserve of synaptic weight growth. So, we can calculate the dynamics of the plasticity potential reserve of synaptic weight growth on the dendrite $d$ using the following equation:

\begin{equation}\label{eq:9}
{\dot{w}}^{pool}_{d} = \sum\limits_{s}{\Delta w^{-}_{ds}} - k^{sat}_{d} \cdot \sum\limits_{s}{\Delta w^{+}_{ds}} + k^{sat}_{soma} \cdot r^{speed} \cdot \Delta w^{pool}_d,
\end{equation}
where $\Delta w^{-}_{ds}$ are the weights returned to the plasticity potential reserve of synaptic weight growth on the dendrite $d$ for the growth of weights at $s$ synapses; $\Delta w^{+}_{ds}$ are the weights consumed from the plasticity potential reserve of synaptic weight growth on the dendrite $d$ for the growth of weights at $s$ synapses; $k^{sat}_{d}$ is the coefficient showing the sufficiency of the plasticity potential reserve of synaptic weight growth to satisfy the amplitude of the growth in all weights on the dendrite $d$; $\Delta w^{pool}_d$ is the demand on the replenishment of the potential reserve of synaptic weight growth $w^{pool}_{d}$ up to maximum amplitude of growth of the sum of weights $w^{res}_{d}$; $r^{speed}$ is the transfer speed coefficient; and $k^{sat}_{soma}$ is the coefficient showing the sufficiency of the plasticity potential reserve of synaptic weight growth to satisfy the amplitude of the growth in all weights on the soma.

The value of ${w}^{pool}_{d}$ depends on the change in the values of the weights at the synapses and is replenished from the pool in the soma. The returned weights $\Delta w^{-}_{ds}$ are found by the following formula:

\begin{equation}\label{eq:10}
\Delta w^{-}_{ds} = - k^{back} \cdot (\dot w_{ds} - w_{ds}) \cdot (1-H(\Delta w_{ds})),
\end{equation}
where $k_{back}$ is the return coefficient of excess changes in synaptic weights to the potential reserve, $k_{back} = 0.2$.

The $\Delta w^{-}_{ds}$ values are always greater than or equal to zero. With a negative $\Delta w_{ds}$, the potential reserve partially (with a coefficient $k_{back}$) will return to ${w}^{pool}_{d}$. If $\Delta w_{ds}$ is positive, $\Delta w^{-}_{ds}$ will be zero. It means that part of the reserve potential will not be lost with a decrease in synaptic weights but will return to the dendrite. 

The consumed weights $\Delta w^{+}_{ds}$ are always zero for $\Delta w_{ds}$ which less than zero:

\begin{equation}\label{eq:11}
\Delta w^{+}_{ds} = (\dot w_{ds} - w_{ds}) \cdot H(\Delta w_{ds}).
\end{equation}

Similarly to the coefficient from Equation~\ref{eq:7}, the coefficient $k^{sat}_{d}$ is in the range from 0 to 1. It will be equal to its maximum value in cases where the current value ${w}^{pool}_{d}$ will be enough to change all weights on the dendrite positively.

\begin{equation}\label{eq:12}
k^{sat}_{d} = 1+\left(\frac{w^{pool}_{d}}{\sum\limits_{s}{\Delta w^{+}_{ds}}} - 1 \right)\cdot H \left( \sum\limits_{s}{\Delta w^{+}_{ds}} - w^{pool}_{d} \right).
\end{equation}

\begin{equation}\label{eq:13}
\Delta w^{pool}_d = (w^{res}_{d} - w^{pool}_{d}) \cdot H(w^{res}_{d} - w^{pool}_{d}),
\end{equation}
where $w^{res}_{d}$ is the maximum amplitude of growth of the sum of weights on the dendrite $d$ in $t$ previous steps.

$\Delta w^{pool}_d$ is always greater than or equal to zero. If $w^{pool}_{d}$ is less than $w^{res}_{d}$, then $\Delta w^{pool}_d$ will be equal to the difference between them, that is, $w^{pool}_{d}$ will be replenished to the level of $w^{res}_{d}$. If $w^{pool}_{d}$ is greater than or equal to $w^{res}_{d}$, then no replenishment occurs and $\Delta w^{pool}_d$ is equal to zero.

The $k^{sat}_{soma}$ coefficient is similar to the $k^{sat}_{d}$, but it checks the sufficiency of the plasticity potential in the soma to replenish the $w^{pool}_{d}$ in all dendrites.

\begin{equation}\label{eq:14}
k^{sat}_{soma} = 1+\left(\frac{w^{pool}_{soma}}{\sum\limits_{d}{\Delta w^{pool}_d}}-1 \right)\cdot H(\sum\limits_{d}{\Delta w^{pool}_d} - w^{pool}_{soma}),
\end{equation}
where $w^{pool}_{soma}$ is the global reserve of the plasticity potential, and it's dynamics is calculated as follows:

\begin{equation}\label{eq:15}
{\dot{w}}^{pool}_{soma} = - k^{sat}_{soma} \cdot r^{speed} \cdot \sum\limits_{d}{\Delta w^{pool}_d}.
\end{equation}

\begin{figure*}
\centering
\includegraphics[width=1.0\textwidth]{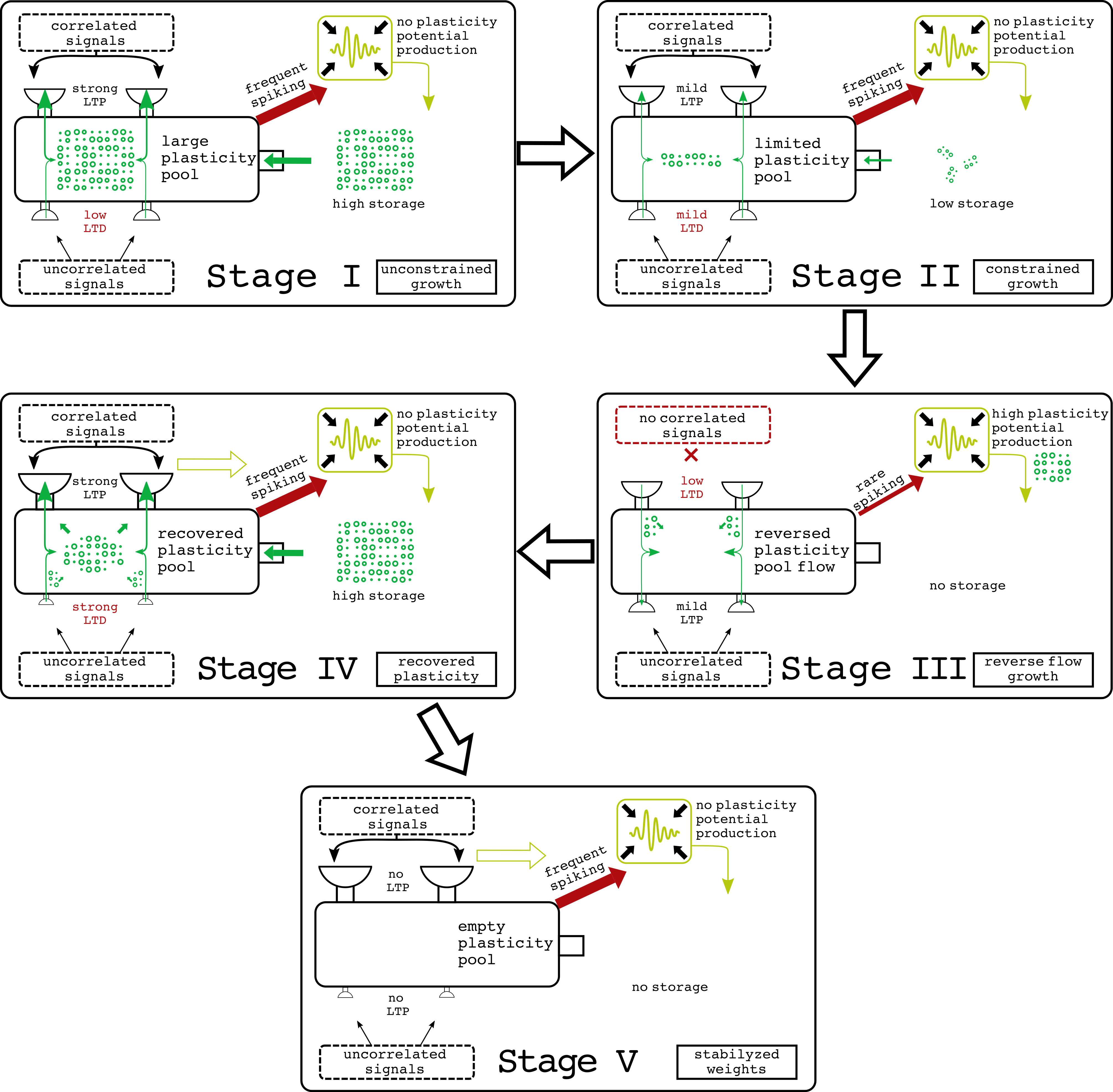}
\caption{One proposed scenario for weight control and dynamics in the model.}
\label{fig:4}
\end{figure*}

$w^{pool}_{soma}$ is decreased by the sum of the values by which $w^{pool}_{d}$ is replenished on all dendrites (see Equation~\ref{eq:9}). The dynamics of the processes described above is schematically shown in Figure~\ref{fig:4}.

Figure~\ref{fig:4} reflects the desired dynamics in the one particular dendrite with weights, restricted by plasticity potential reserve. Before the process of stimulation, all weights on the dendrite are equal. However, part of them stimulated with random independent noise and another part with correlated input signals. Therefore, it will increase the neuron's spiking activity (Figure~\ref{fig:4}, Stage I). As there is a lot of weight growth potential in the dendrite, nothing prevents weights from growing according to the STDP rule, and some synapses with correlated activity will initiate a strong LTP, and those with uncorrelated will experience mild LTD. In response to the increased firing rate, the control system will reduce plasticity materials release. It will lead to the next stage (Figure~\ref{fig:4}, Stage II). At this stage, a limited reserve of plasticity potential materials constrains the weight growth. It will lead to the growth of correlated stimulated synapses at the expense of shrinking others. 

If the correlated input stops (Figure~\ref{fig:4}, Stage III), synapses with at least some uncorrelated input will grow at the expense of decreasing previously outgrown synapses. As the neuron will start only receiving random uncorrelated noise, this will decrease spiking frequency, and the control system will release more plasticity potential materials. It will be delivered to dendrite, and after the recovery of correlated input, synapses receiving that input will grow to the point when they will initiate action potentials without any random co-stimulation (Figure~\ref{fig:4}, Stage IV). It will lead to strong divergence in synaptic wights of correlated and uncorrelated inputs. A strong reaction to correlated inputs will initiate strong firing output, leading to the interruption of plasticity potential materials reserve. If all plasticity potential materials and input signals persist on the same level, the synaptic weights will stabilize due to SISS at the dendrite (Figure~\ref{fig:4}, Stage V).

The described weight dynamics restricted by plasticity potential materials reserve should lead to higher weight divergence between correlated and uncorrelated inputs and eventually settle weights on some 'saddle points.'

Note that we can regulate the filtering of synaptic noise by manipulating with target firing rate. If the target frequency $\theta_{target}$ is high, the neuron will act as a basic STDP and will allow synapses to develop easily. While at low $\theta_{target}$, it will let to develop only synapses with strongly correlated and limited frequencies. 

So we described all necessary parts of our proposed framework, such as the basic plasticity rule, the spiking behavior, the dynamics of synaptic plasticity potential storage, and the overall paradigm of its regulation. Next, we introduce the particular control approaches that we use in the present paper to regulate plasticity potential storage in the neuron and its dendrites.

Thus, we use the following algorithm of the neuron plasticity regulation model (Algorithm~\ref{al:1}).

\begin{algorithm}\label{al:1}
\caption{Neuron plasticity regulation model}
\begin{algorithmic}[1]
\REQUIRE $I_{ds}$
\STATE Calculate integrated weighted input $I$ to the neuron (Equation~\ref{eq:1})
\STATE Generate the integrated output $O$ from the spiking model
\STATE Update $\Delta w^{stdp}_{ds}$ (Equations~\ref{eq:a13}--\ref{eq:a14})
\STATE Recalculate $\Delta w^{stdp}_{ds}$ (Equations~\ref{eq:6}--\ref{eq:7}) based on the availability of plasticity potential in each dendrite ($w^{pool}_{d}$) and finally update $w_{ds}$ (Equations~\ref{eq:5})
\STATE Set the new state of the plasticity potential reserve $w^{pool}_{soma}$ (particular implementations of control approaches are described in Section~\ref{subsec:32}) with respect to firing frequency $\theta_{real}$ control
\STATE Update $w^{pool}_{d}$ based on the new $w^{pool}_{soma}$ (Equations~\ref{eq:9}--\ref{eq:15})
\RETURN New step
\end{algorithmic}
\end{algorithm}

At first, in the neuron, synapses receive inputs in the form of '0' and '1', where '0' means that no signal is received, and '1' -- is the received signal. Inputs $I_{ds}$ are multiplied by the weights $w_{ds}$, and then scaled and directed to the spiking model. Based on the timing between inputs and action potential generation ($O$), we calculate the update for weights, $\Delta w^{stdp}_{ds}$. After that, the weight growth is regulated by a control mechanism. In our case, we test and compare three different kinds of control algorithms, described in Sections~\ref{subsubsec:321}--\ref{subsubsec:323}:

\begin{enumerate}
\item firing frequency-dependent allocation model (FFDA),
\item echo state network control (ESN) \cite{ESN},
\item backward calculation of plasticity potential reserve demand (PPD).
\end{enumerate}

Plasticity potential reserve ($w^{pool}_{soma}$) is assigned by one of three algorithms. Plasticity potential reserve in dendrites $w^{pool}_{d}$ is then refilled from $w^{pool}_{soma}$. The $w^{pool}_{d}$ determines the ability of weights in a particular dendrite to grow according to STDP weight update, and we get the final weights after the one loop cycle.

Therefore, this algorithm of neuron plasticity regulation describes the processes in the abstract neuron, including the change in the synaptic weights and control of the plasticity potential reserve. Below we describe three $w^{pool}_{soma}$ control algorithms.

\subsubsection{FFDA}\label{subsubsec:321}

FFDA represents the restriction mechanism when if the firing frequency $\theta_{real}$ is less than the target frequency $\theta_{target}$, then the plasticity potential reserve in soma $w^{pool}_{soma}$ is assigned according to the difference between these frequencies (Equation~\ref{eq:16}).

\begin{equation}\label{eq:16}
w^{pool}_{soma} = 
\begin{cases}
\begin{alignedat}{3}
\sum\limits_{d}{w}^{res}_{d} \cdot (\theta_{target} - \theta_{real}), 
& \text{ if } \theta_{real} < \theta_{target} \\
0, 
& \text{ otherwise} 
\end{alignedat} 
\end{cases}
\end{equation}

When $\theta_{real}$ is not enough to reach the target frequency $\theta_{target}$, then we begin to add the plasticity potential to prevent $\theta_{real}$ from falling below the target. 

\subsubsection{ESN}\label{subsubsec:322}

The disadvantage of the approach from Section \ref{subsubsec:321} is a lack of adaptability and reliance on the history with backward window, which leads to a delay in calculating the required amount of $w^{pool}_{soma}$. Hence, it cannot be applied to complex tasks and networks where greater accuracy is required. For more complex tasks, we need to control $w^{pool}_{soma}$ by some adaptive algorithm. It might be either the black box kind of control system or the direct numeric calculation of optimal $w^{pool}_{soma}$. Here, we provide the approach to regulating weight potential reserve by ESN \cite{ESN} as an approach with universal computing capabilities inspired by the cellular GRN/PPI networks. We train it with reinforcement learning as it is hard to calculate the exact learning labeling for training in our setup. 

Genetic regulation is a nonlinear dynamic information processing that occurs in the network structure and has a complex interaction graph consisting of the work of GRN and protein interactions. Modeling of genetic regulatory networks of cells using ESN has been proposed in \cite{4}. ESN in the task of reinforcement learning was proposed in \cite{Szita} and consisted of adapted SARSA Q-learning, where ESN is used to approximate the action-state pairs. We also propose to use the ESN for the creation of an adapted $Q$-learning system \cite{QL} for dynamic optimization of the neuron parameters. (Table 1) 

\begin{table}\label{tab:1}
\caption{Proposed $Q$-learning parameters for ESN}
\begin{tabular*}{\hsize}{@{\extracolsep{\fill}}lll@{}}
\hline
Parameters & Values \\
\hline
State      &   $w^{pool}_{d}$, $w_{ds}$, $\theta_{real}$ \\
Actions    &   $w^{pool}_{soma}=0$ or $w^{pool}_{soma}=\sum\limits_{d}{w}^{res}_{d}$ \\
$Q$-values &   Divergence from $\theta_{target}$ level \\
\hline
\end{tabular*}
\end{table}

Intraneural ESN controls $w^{pool}_{soma}$ during the $Q$-learning, thus, constraining the growth of synaptic weights. The neuron is trying to optimize the firing frequency level $\theta_{real}$. The level of the plasticity potential ($w^{pool}_{soma}$) is controlled by ESN giving two options of actions: provide the $w^{pool}_{soma}$ to dendrites or not. Neuronal ESN reinforcement learning controls internal values.

Our ESN consists of two input layers, the reservoir size is 25, the contraction coefficient is 0.9, the density is 1.0, and the scale input coefficient is 1.0, $\gamma=0.4$, $\epsilon=0.05$, and it gradually decreases to 0.005 by the last step.

For the ESN control model, we use reinforcement learning with negative reward, so when the ESN does not reach the target value, it receives a penalty proportional to the deviation from the target. The reward for ESN is calculated as:

\begin{equation}\label{eq:17}
R = - |\theta_{target}-\theta_{real}|\cdot k^{rew},
\end{equation}
where $k^{rew}$ is the reward scaling coefficient, here, $k^{rew}=10$.

\subsubsection{PPD}\label{subsubsec:323}

If we consider that future inputs to the neurons might be sampled based on the known history of previous inputs, then we may assume the future STDP weight change and demand across all the synapses and calculate the exact amount of $w^{pool}_{soma}$ that should be injected into the model to shift the firing frequency of a neuron towards the desired value $\theta_{target}$. It cannot be done with ideal precision, but if we make assumptions and calculations for each step, it should drive the frequency towards an optimal firing frequency state.

We propose an algorithm to control the firing frequency by calculating the global reserve of the plasticity potential $w^{pool}_{soma}$ and correcting it. (Algorithm~\ref{al:2}) 

\begin{algorithm}\label{al:2}
\caption{Control of firing frequency based on $w^{pool}_{soma}$ allocation by direct demand calculation}
\begin{algorithmic}[1]
\REQUIRE $I_{ds}$, $\theta_{target}$
\STATE Calculate $w^{min}_{ds}$ (if $w^{pool}_{d}=0$) and $w^{max}_{ds}$ (if $w^{pool}_{d}=w^{res}_{d}$)
\STATE Calculate $\theta^{t-wind}_{input}$
\STATE Calculate $I^{t+wind}_{ds}$ using sampling $I^{t-wind}_{ds}$ with $\theta^{t-wind}_{input}$ and Bernoulli distribution
\STATE Calculate $I^{t+wind}_{ds}\cdot w^{min}_{ds}$ and $I^{t+wind}_{ds}\cdot w^{max}_{ds}$
\STATE Get $\theta_{min}$ and $\theta_{max}$ from Equations~\ref{eq:a1}--\ref{eq:a2}
\STATE Compare with $\theta_{target}$:
    \IF{$\theta_{target}<\theta_{min}$}
        \STATE Set $w^{pool}_{soma}=0$
    \ELSIF{$\theta_{target}>\theta_{max}$}
        \STATE Set $w^{pool}_{soma}=\sum\limits_{d}{w}^{res}_{d}$
    \ELSIF{$\theta_{min}\leq \theta_{target}\leq \theta_{max}$}
        \STATE Calculate $w^{pool}_{soma}$ using Equations~\ref{eq:18}--\ref{eq:22}
    \ENDIF
\RETURN $w^{pool}_{soma}$
\end{algorithmic}
\end{algorithm}

For this, the required frequency $\theta_{target}$ is set. Then, the theoretically possible minimum $w^{min}_{ds}$ (if $w^{pool}_{d}$ is equal to zero) and maximum $w^{max}_{ds}$ (if $w^{pool}_{d}$ is maximum possible) values of the weights are calculated. After that, the estimated inputs $I^{t+wind}_{ds}$ for a given forward window $wind$ are calculated: the frequency of inputs $\theta^{t-wind}_{input}$ for the $(t-wind)$ time steps (backward window) is calculated and the inputs $I^{t-wind}_{ds}$ or the $(t-wind)$ time steps are sampled with the received frequency $\theta^{t-wind}_{input}$ and Bernoulli distribution. The estimated inputs $I^{t+wind}_{ds}$ are multiplied by the corresponding weights and put into the spiking model (from Equations~\ref{eq:a1}--\ref{eq:a2}), resulting in the estimated $\theta_{min}$ and $\theta_{max}$ frequencies. 

We compare the obtained frequencies with a given frequency $\theta_{target}$: if $\theta_{target}$ is less than $\theta_{min}$, then we zero the $w^{pool}_{soma}$, if it is greater than $\theta_{max}$, then we take $w^{pool}_{soma}$ equal to the maximum possible value $w^{res}_{d}$, if it is in the interval between them, then we calculate $w^{pool}_{soma}$ using backward calculation (Equations~\ref{eq:18}--\ref{eq:22}):

\begin{equation}\label{eq:18}
\begin{aligned}
{\dot{w}}^{pool}_{soma} = {} & \sum\limits_{d}\left(\frac{\sum\limits_{d}{\Delta w^{pool}_{d}}\cdot ({\dot{w}}^{pool}_{d}- w^{pool}_{d} + \sum\limits_{s}{\Delta w^{-}_{ds}})}{\Delta w^{pool}_{d} \cdot r^{speed} \cdot H^{wp}_{soma}}\right) \\
      & + \sum\limits_{d}\left(\frac{\sum\limits_{d}{\Delta w^{pool}_{d}} \cdot k^{sat}_{d} \cdot \sum\limits_{s}{\Delta w^{+}_{ds}}}{\Delta w^{pool}_{d} \cdot r^{speed} \cdot H^{wp}_{soma}}\right) \\
      & - \sum\limits_{d}\left(\frac{\sum\limits_{d}{\Delta w^{pool}_{d}}\cdot (1-H^{wp}_{soma})}{H^{wp}_{soma}}\right),
\end{aligned}
\end{equation}
where
\begin{equation}\label{eq:19}
H^{wp}_{soma} = H(\sum\limits_{d}{\Delta w^{pool}_{d}} - w^{pool}_{soma}),
\end{equation}

\begin{equation}\label{eq:20}
\begin{aligned}
{\dot{w}}^{pool}_{d} = {} & \sum\limits_{s}\left(\frac{\sum\limits_{s}{\Delta w^{stdp+}_{ds}}\cdot ({\dot{w}}^{new}_{ds}-k^{siss}_{d}\cdot w_{ds})}{\Delta w^{stdp}_{ds}\cdot H^{wp}_{d}}\right) \\
      & - \sum\limits_{s}\left(\frac{\sum\limits_{s}{{\Delta w^{stdp+}_{ds}}\cdot k^{siss}_{d}\cdot (1-H^{wp}_{d})}}{H^{wp}_{d}}\right),
\end{aligned}
\end{equation}
where
\begin{equation}\label{eq:21}
H^{wp}_{d} = H\left( \sum\limits_{s}{\Delta w^{stdp+}_{ds}} - \frac{w^{pool}_{d}}{k^{siss}_{d}}\right)\cdot H(\Delta w^{stdp}_{ds}),
\end{equation}

\begin{equation}\label{eq:22}
{\dot{w}}^{new}_{ds} = w^{min}_{ds}+(w^{max}_{ds}-w^{min}_{ds})\cdot \frac{(\theta_{target}-\theta_{min})}{(\theta_{max}-\theta_{min})},
\end{equation}
where $w^{min}_{ds}$ and $w^{max}_{ds}$ are the minimum and the maximum value of the synaptic weights; $\theta_{target}$, $\theta_{min}$, and $\theta_{max}$ denote the target, the minimum and the maximum firing frequency.

Above, we provided four kinds of approaches of restricting weight outgrowth based on keeping firing frequency goal. These approaches should be tested for reaction on different synaptic input simulations resembling some regimes that might emerge in spiking neural networks. We will compare these models to the performance of the basic STDP weight update rule. The following section covers such testing.  

\section{Results}\label{sec:4}

To explore the approaches described above, we run a series of experiments comparing the following weight outgrowth restriction models (as well as the basic STDP weight update rule model): basic STDP weight update rule (Equations~\ref{eq:a13}--\ref{eq:a14}), HSS, FFDA, ESN, and PPD.

In all experiments, we pass the same inputs for all types of models. In various experiments, depending on the task, either low-frequency ($\theta_{input}=0.2$) or high-frequency ($\theta_{input}$\allowbreak = 0.5) inputs are used. Input frequencies are normalized into the range [0,1]. Also, there are one neuron, three dendrites, and six synapses used for all experiments. Each experiment takes 2400 steps, and 100 such tests are performed for each task. We average the results for 100 tests for display on figures, excluding figures for inputs and spiking.

Two setups, except for the experiments with correlated inputs, are compared: with SISS in the model or not. The presented modeling results contain such indicators as: input, input frequency, spiking $O$, spiking frequency $\theta_{real}$, $w_{ds}$ and $w^{pool}_d$ for all dendrites, $w^{pool}_{soma}$. Below, we present the results of experiments with 4 different protocols.

The experimental protocols for comparing the models with different approaches described above are following:

\begin{enumerate}
\item constant stimulation;
\item discrete neuron-wise high-frequency perturbations;
\item discrete dendrite-specific high-frequency perturbations;
\item experiments with correlated inputs.
\end{enumerate}

\subsection{Constant stimulation}\label{subsec:41} 

\begin{figure*}
\centering
\includegraphics[width=1.0\textwidth]{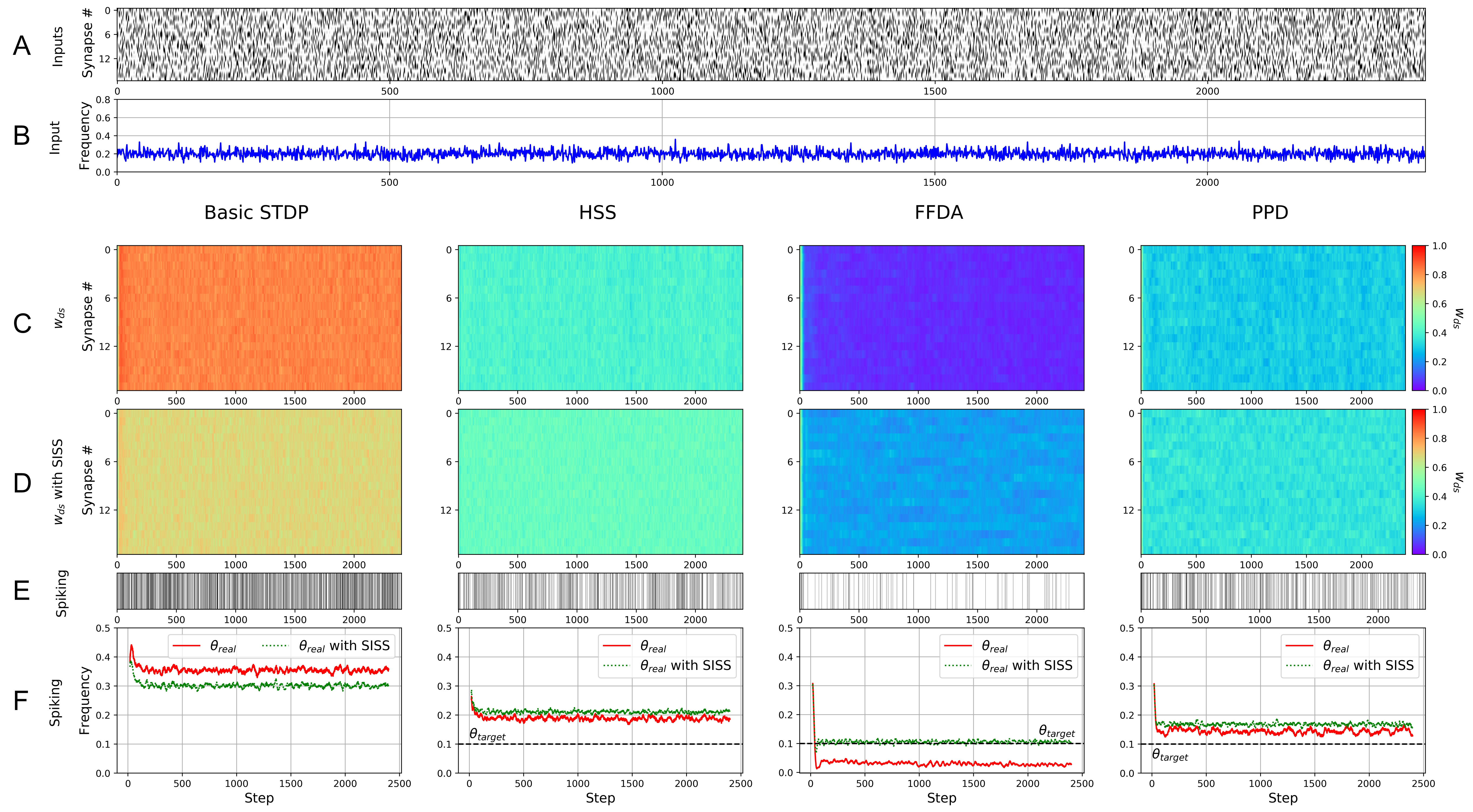}
\caption{Constant stimulation experiment. All parameters in graphs in the figure are calculated for 2400 steps each of the basic STDP, HSS, FFDA, and PPD models, respectively, and averaged for all synapses and dendrites over 100 tests for each model (except (A), (E), are the non-averaged examples from the last test). (A) Inputs arrive at all synapses with constant frequency ($\theta_{input}=0.2$) equal for all models. (B) Input frequency. (C) Dynamics of the synaptic weights $w_{ds}$ without SISS,  where the color indicates the $w_{ds}$ value. (D) Dynamics of the synaptic weights $w_{ds}$ with SISS, where the color indicates the $w_{ds}$ value. (E) Action potential generation (spiking) $O$ of the neuron ('0' and '1' values) during the work of the models. (F) Spiking frequency $\theta_{real}$ with and without SISS and level of target frequency $\theta_{target}$.}
\label{fig:5}
\end{figure*}

In the present paper, we test the dynamics of different kinds of frequency homeostasis-based algorithms of weight plasticity control. First, we test the saturation of weights. All synapses of all dendrites receive inputs with a constant frequency ($\theta_{input}=0.2$). Thus, there is constant stimulation of the neuron. For HSS, FFDA, and PPD models, the $\theta_{target}$ is 0.1, and for basic STDP, the $\theta_{target}$ is not assigned. It is the simplest experiment that allows us to study the system's performance in a stable environment without disturbances. (Figure~\ref{fig:5})

Figure~\ref{fig:5} presents the basic STDP, HSS, and two approaches to the weight regulation based on plasticity potential reserve control (FFDA and PPD). It also shows tests with and without SISS. In Figure~\ref{fig:5}C, the heatmap for all synaptic weights for all the algorithms provided. It can be seen that with pure STDP, synaptic weights do grow to high values. On the other hand, SISS keeps for the basic STDP those values lower because it constantly counteracts against the weights' extensive growth towards the goal point of dendritic weights $\sum\limits_{s}w^{ideal}_{ds}$. 

For non-SISS results, the HSS model maintains a stable $\theta_{real}$, but does not reach $\theta_{target}$. The FFDA model sharply reduces the values of the weights, as a result of which its $\theta_{real}$ becomes close to zero. After that, the model cannot restore the spike rate to the $\theta_{target}$ level. The PPD model is closest to all the previous ones approaching $\theta_{target}$. As a result of using the SISS for HSS, FFDA, and PPD, their $\theta_{real}$ stabilize and show smaller fluctuations in the spiking frequency. At the same time, the FFDA model with SISS manages to achieve $\theta_{real}$ by curbing the weight reduction, thanks to SISS. For the models HSS, FFDA, and PPD, SISS drives the frequency slightly above non-SISS. It constantly drives the weights to 0.5, while their optimal values to keep the low firing rate $\theta_{target}$ of 0.1 are lower than $w^{ideal}_{ds}$. To further test the algorithm, more complex experiments are needed.

\subsection{Discrete neuron-wise high-frequency perturbations}\label{subsec:42}

\begin{figure*}
\centering
\includegraphics[width=1.0\textwidth]{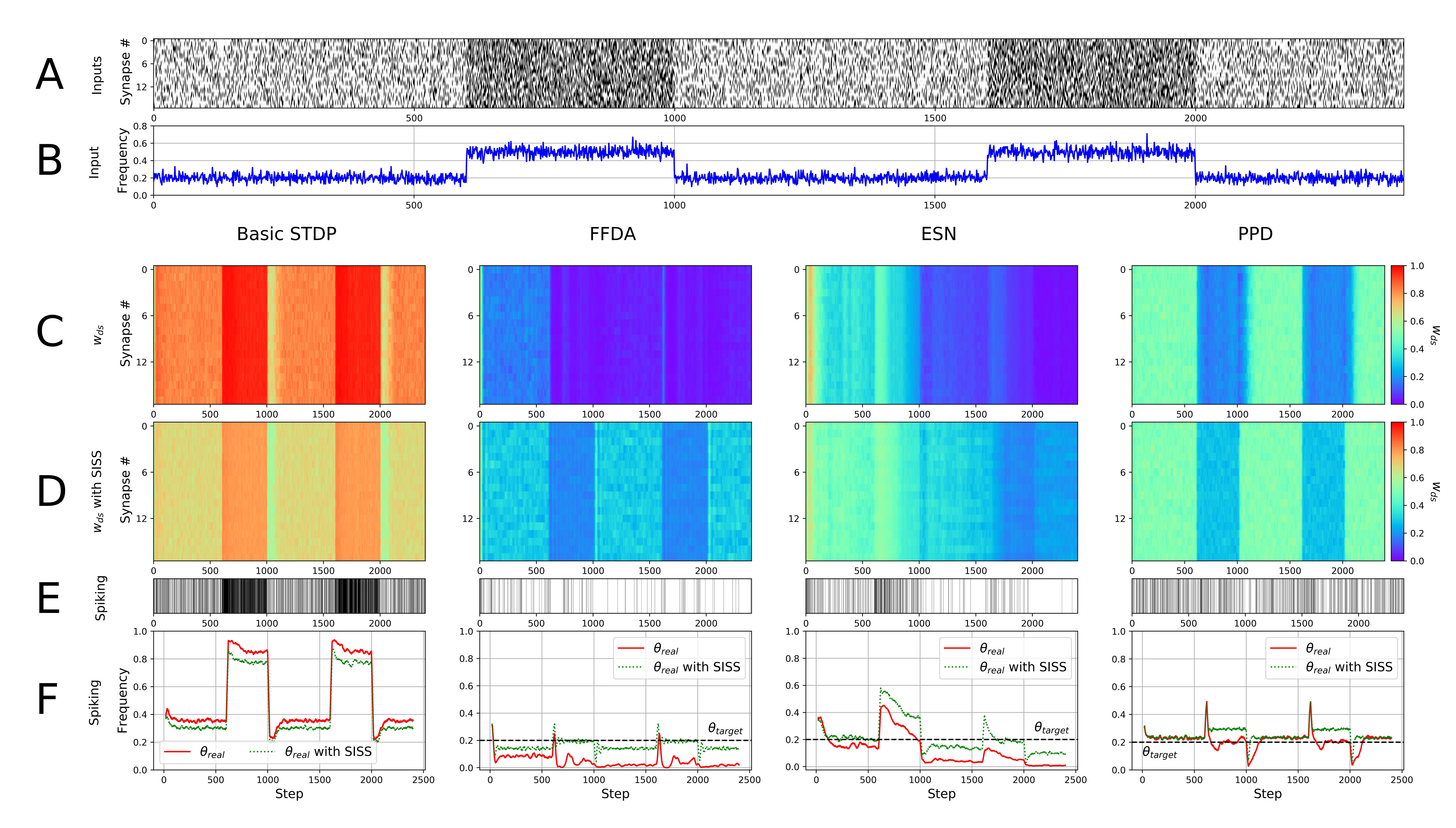}
\caption{The results of the experiment with discrete neuron-wise high-frequency perturbations. All parameters in graphs in the figure are calculated for 2400 steps each of the basic STDP, FFDA, ESN, and PPD models, respectively, and averaged for all synapses and dendrites over 100 tests for each model (except (A), (E), are the non-averaged examples from the last test). (A) Inputs with frequency $\theta_{input}=0.2$ equal for all models arrive at all synapses, but from 600 to 1000 steps and from 1600 to 2000 steps the frequency of inputs changes to $\theta_{input}=0.5$. (B) Input frequency. (C) Dynamics of the synaptic weights $w_{ds}$ without SISS,  where the color indicates the $w_{ds}$ value. (D) Dynamics of the synaptic weights $w_{ds}$ with SISS, where the color indicates the $w_{ds}$ value. (E) Action potential generation (spiking) $O$ of the neuron ('0' and '1' values) during the work of the models. (F) Spiking frequency $\theta_{real}$ with and without SISS and level of target frequency $\theta_{target}$.}
\label{fig:6}
\end{figure*}

To test the dynamics of weight control reaction to different inputs, experiment with discrete neuron-wise high-frequency perturbations are carried out, where all synapses of all dendrites of a neuron continually receive low-frequency inputs ($\theta_{input}=0.2$), but from 600 to 1000 steps and from 1600 to 2000 steps the frequency of inputs changes to $\theta_{input}=0.5$. For FFDA, ESN, and PPD models, the $\theta_{target}$ is 0.2, and for basic STDP, the $\theta_{target}$ is not assigned. Thus, at the moments of frequency change, high-freq-\allowbreak uency inputs received by the model can be compared with disturbances in the input signal of the neuron and impose an additional need on the system to regulate this process. We test basic STDP and three approaches to plasticity potential reserve control to find out how efficient are algorithms proposed in the paper above. (Figure~\ref{fig:6})

In Figure~\ref{fig:6} it can be seen that in contrast to STDP, all weight restricting approaches lead to the decrease of more frequent input weights. It is because weight increase drives neurons out of the firing frequency homeostatic state. We can also see that, with no SISS, the FFDA and ESN-based approaches fail to stabilize firing frequency and slowly decrease it, but despite this fact, they react to firing frequency jumps. The FFDA and ESN models overcompensate the weight limitations. Only PPD is correct in control. SISS upregulation stabilizes the control of $\theta_{target}$.

This section proves that the plasticity potential reserve control approaches do function in practice, and we may choose one for future testing and practice use. Given the complexity of tuning, control quality, and computing cost, ESN for control is unfavorable. However, the neuron model for this paper includes different dendrites. In the following section, we test how proposed algorithms will act in a more nonlinear setup.

\subsection{Discrete dendrite-specific high-frequency perturbations}\label{subsec:43}

It is necessary to provide dendrites with different input intensities to test the dynamics of weights between different dendrites. That is why the signals are applied selectively to different synapses.

\begin{figure*}
\centering
\includegraphics[width=1.0\textwidth]{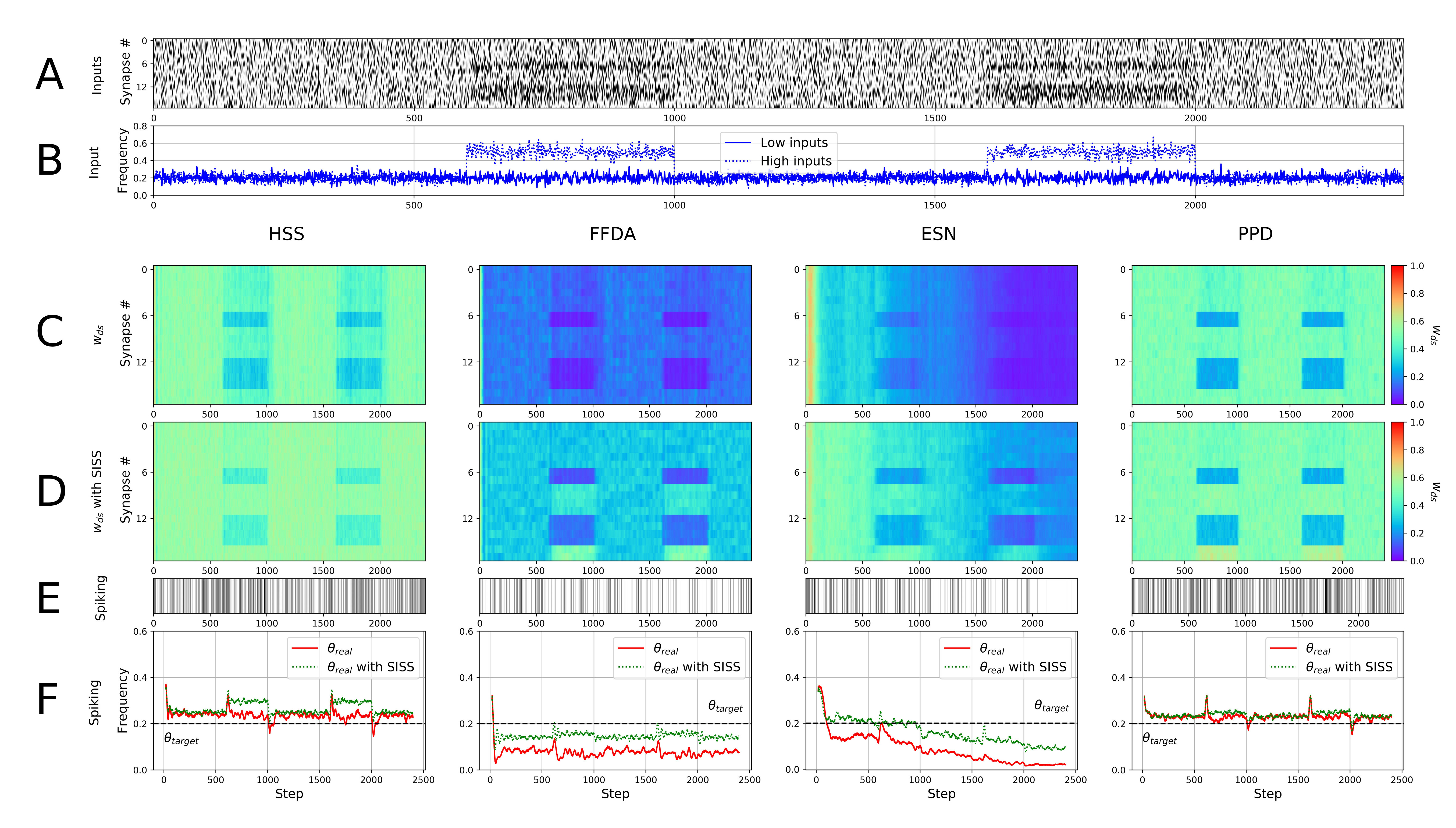}
\caption{The results of the experiment with discrete dendrite-specific high-frequency perturbations. All parameters in graphs in the figure are calculated for 2400 steps each of the HSS, FFDA, ESN, and PPD models, respectively, and averaged for all synapses and dendrites over 100 tests for each model (except (A), (E), which are the non-averaged examples from the last test). (A) Inputs with frequency $\theta_{input}=0.2$ equal for all models arrive at all synapses of the 1st dendrite, 4 synapses of the 2nd dendrite, and 2 synapses of the 3rd dendrite, while at 2 synapses of the 2nd dendrite and 4 synapses of the 3rd dendrite from 600 to 1000 steps and from 1600 to 2000 steps arrive high-frequency inputs ($\theta_{input}=0.5$). (B) Input frequency, where 'Low' and 'High' are low-frequency and high-frequency inputs, respectively. (C) Dynamics of the synaptic weights $w_{ds}$ without SISS, where the color indicates the $w_{ds}$ value. (D) Dynamics of the synaptic weights $w_{ds}$ with SISS, where the color indicates the $w_{ds}$ value. (E) Action potential generation (spiking) $O$ of the neuron ('0' and '1' values) during the work of the models. (F) Spiking frequency $\theta_{real}$ with and without SISS and level of target frequency $\theta_{target}$.}
\label{fig:7}
\end{figure*}

So, we implement the experiment with discrete dendrite-specific high-freq-\allowbreak uency perturbations as follows: all synapses of the first dendrite, four synapses of the second dendrite, and two synapses of the third dendrite receive low-frequency inputs ($\theta_{input}=0.2$), while two synapses of the second dendrite and four synapses of the third dendrite receive high-frequency inputs from 600 to 1000 steps and from 1600 to 2000 steps ($\theta_{input}=0.5$). For all models, the $\theta_{target}$ is equal to 0.2. All the rest of the time, except for these steps, the last ones also receive low-frequency inputs. This experiment repeats the previous (the second) one, but the input frequency varies depending on the dendrites. (Figure~\ref{fig:7})

\begin{figure*}
\centering
\includegraphics[width=1.0\textwidth]{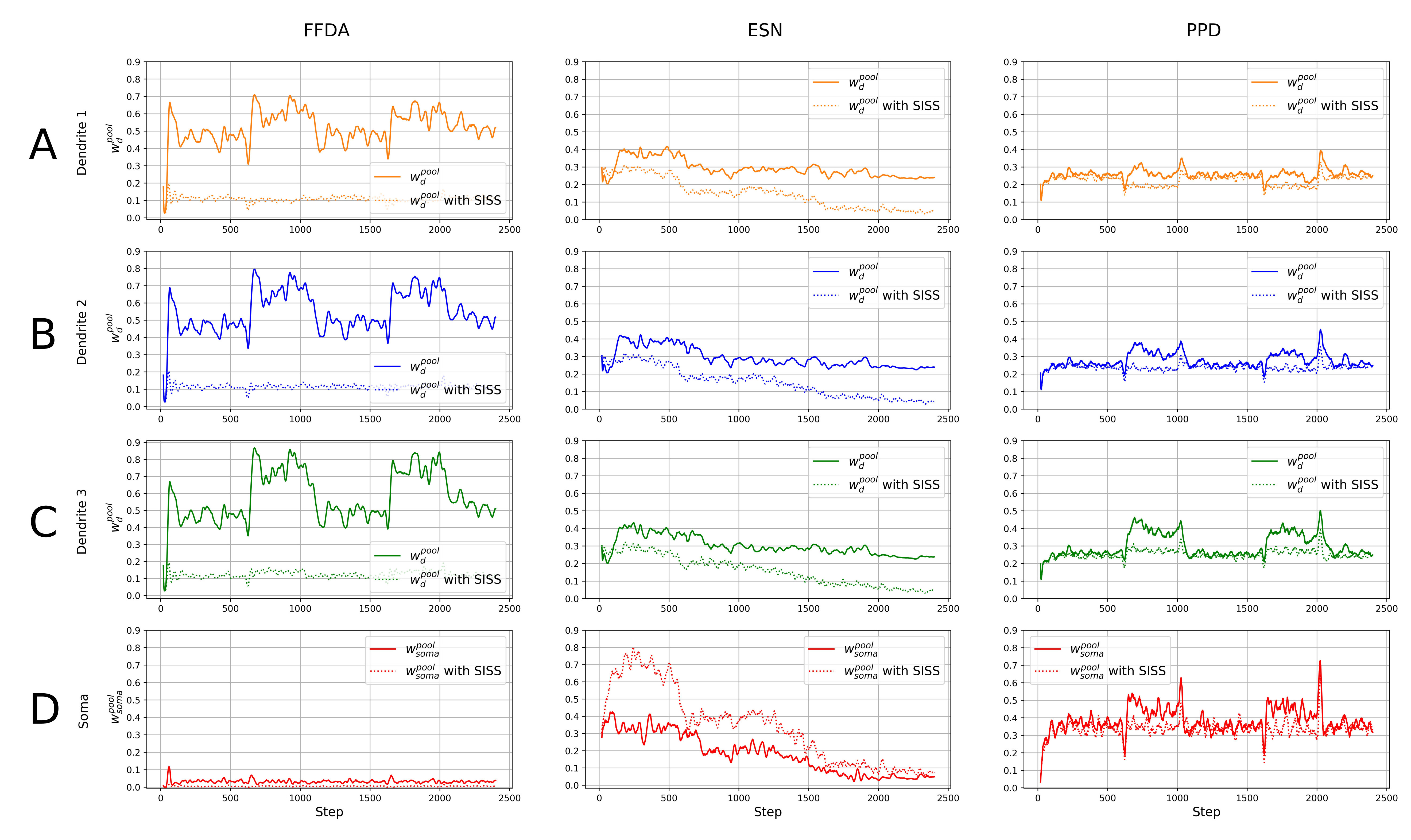}
\caption{Comparison of the plasticity potential reserve in dendrites for each dendrite and the global reserve of the plasticity potential of three models: FFDA, ESN, PPD. All parameters in graphs in the figure are calculated for 2400 steps each of the FFDA, ESN, and PPD models, respectively, and averaged for all synapses and dendrites over 100 tests for each model. (A)-(C) Dynamics of the plasticity potential reserve in dendrites $w^{pool}_{d}$ with and without SISS on the 1st, 2nd and 3rd dendrites, respectively. (D) Dynamics of global reserve of the plasticity potential $w^{pool}_{soma}$ with and without SISS.}
\label{fig:8}
\end{figure*}

Figure~\ref{fig:8} shows the changes in the plasticity potential reserve in dendrites $w^{pool}_d$ for each dendrite and the global reserve of the plasticity potential $w^{pool}_{soma}$ of three models -- FFDA, ESN, PPD. Furthermore, each graph compares two types of $w^{pool}_d$ -- with SISS and without it. 

The HSS reduces the weights at the synapses to approach the target frequency. (Figure~\ref{fig:7}) It manages to adhere to the target frequency during periods with high-frequency inputs. At the same time, it can be seen that the weights change simultaneously on all dendrites, including on the first dendrite; six of its synapses do not receive a high-frequency input. The SISS in HSS does not allow to reduce the weights during these intervals to the desired degree, and the HSS itself cannot redistribute the weights at synapses. As a result, the spiking frequency in these intervals rises above the target.

In both cases (with and without SISS), the FFDA decreases in weights at all synapses, but with SISS, they do not fall so low because it restrains them. Moreover, when there is a SISS, a redistribution of weights occurs to an insignificant extent. Synapses with high-frequency inputs decrease weights; due to this, weights increase at synapses with low frequencies located on the same dendrites. The $\theta_{real}$ is below the $\theta_{target}$, while, due to the SISS, the $\theta_{real}$ for this case is closer to the $\theta_{target}$. The Figure~\ref{fig:8}D of $w^{pool}_{soma}$ shows that in cases with SISS, a decrease in weights is suppressed in comparison with the model without SISS. Therefore, it can be seen that $w^{pool}_d$ for the case with SISS arrive into dendrites very few. For the FFDA without SISS, $w^{pool}_{soma}$ was also restricted. However, $w^{pool}_{d}$ has sufficient reserve, which is not consumed due to rare spikes.

The weights at the synapses rise at the first steps in the ESN. Due to this, the $\theta_{real}$ is higher than the $\theta_{target}$. After that, we train the ESN to reduce the weights, and by the time the first interval with high-frequency inputs begins, its spiking frequency is close to the target. After the end of this interval, the frequency of inputs decreases, and at the same time, the spiking frequency decreases. However, the model continues to adhere to reducing the weights, showing its inability to overfit when the task conditions change. For the case without SISS, this strategy leads to a significant reduction in weights that spiking almost stops after 2000 steps. In the ESN with SISS, the process of weight reduction is more restrained, but the strategy itself does not change. Figure~\ref{fig:8} clearly shows the strategy of decreasing $w^{pool}_{soma}$ and the subsequent decrease of  $w^{pool}_{d}$. At the same time, we see that for the case with SISS before the first interval with high-frequency inputs, the ESN successfully learned to increase $w^{pool}_{soma}$. However, after that, it could not return to this strategy.

In the PPD, during intervals of high-frequency inputs, the weights decrease to a greater extent at active synapses, while in the case of SISS, a more pronounced redistribution of weights occurs on the second and third dendrites. The PPD shows its effectiveness in achieving the target frequency, while with the SISS, there is no significant discrepancy between the $\theta_{real}$ and the $\theta_{target}$. We see a stable strategy of setting $w^{pool}_{soma}$ for all intervals with low-frequency inputs and boosting $w^{pool}_{soma}$ for high-frequency intervals. For the PPD with SISS, the $w^{pool}_{soma}$ generation is more stable throughout the entire experiment.

It could be concluded from the current experiment that homeostatic frequency control approaches have filtered disturbing high-frequency inputs. The FFDA and ESN can successfully signal discrimination only of the presence of SISS, while HSS and PPD can adaptively filter with no additional mechanisms. The PPD not only filtered high-frequency input noise but also heterosynaptically upscaled inputs with low frequencies in the presence of SISS. PPD with SISS performs the best as a self-learning filter on a particular task.

\subsection{Experiments with correlated inputs}\label{subsec:44} 

Local learning rules in spiking neural networks, such as STDP, are desired to find correlations in the data. Indeed the original Hebb's postulate stated: "fire together - wire together" means that it should correlate between the presynaptic stimulus and postsynaptic output. That is why it is most important to find out how STDP and its constrained versions (HSS and plasticity reserve-based approach) react on correlated signals. We also compare the reactions to the intense input with high frequency and high-correlated input. As we have seen above, high-frequency input leads to weight degradation. However, the coincidental arrival rate of frequent and correlated inputs should be similar, and this points that it is essential to study such an interaction. Also, it is crucial for brain and artificial neural networks to keep a good signal-to-noise ratio when the stimulus is easily distinguishable from the background. This article is devoted to the frequency homeostatic models as it is essential for neural network dynamics to keep the mean activation stable on the criticality edge. Below, we will inspect the plasticity control approaches with correlated stimulus protocol to test the models to respond to changes in the input frequency and distinguish between correlated inputs.

\begin{figure*}
\centering
\includegraphics[width=1.0\textwidth]{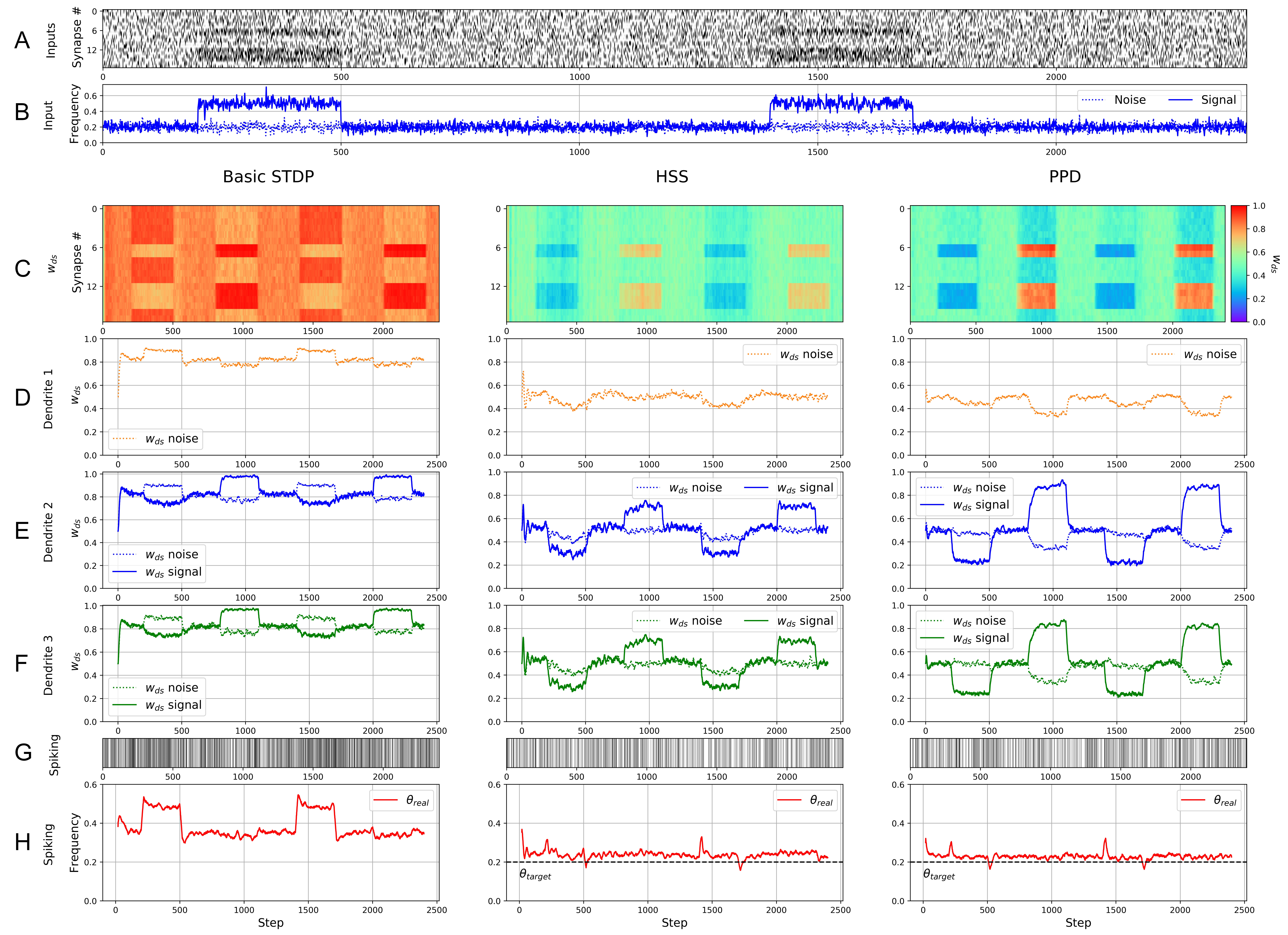}
\caption{The results of the experiments with correlated inputs. All parameters in graphs in the figure are calculated for 2400 steps each of the basic STDP, HSS, and PPD models, respectively, and averaged for all synapses and dendrites over 100 tests for each model (except (A), (G), which are the non-averaged examples from the last test). (A) Inputs with frequency $\theta_{input}=0.2$ arrive at all synapses of the 1st dendrite, 4 synapses of the 2nd dendrite, 2 synapses of the 3rd dendrite. The rest of the synapses (2 synapses of the 2nd dendrite and 4 synapses of the 3rd dendrite) receive the following inputs: from 200 to 500 steps and from 1400 to 1700 steps are high-frequency inputs ($\theta_{input}=0.5$), and from 800 to 1100 steps and from 2000 to 2300 steps correlated signals are included with $\theta_{input}=0.2$. (B) Input frequency, where the only noise is fed ('Noise') or contains correlated signals ('Signal'). (C) Dynamics of the synaptic weights $w_{ds}$ for all synapses and dendrites,  where the color indicates the $w_{ds}$ value. (D)-(F) Dynamics of the synaptic weights $w_{ds}$ on the 1st, 2nd and 3rd dendrites respectively, '$w_{ds}$ noise' and '$w_{ds}$ signal' are synaptic weights to which only noise is fed ('$w_{ds}$ noise') or contains correlated signals ('$w_{ds}$ signal'). (G) Action potential generation (spiking) $O$ of the neuron ('0' and '1' values) during the work of the models. (H) Spiking frequency $\theta_{real}$ and level of target frequency $\theta_{target}$.}
\label{fig:9}
\end{figure*}

For this task, we use correlated inputs. We assume that some of the inputs are correlated. To create such inputs, we generate an initial 'mask' input. Then, we select synapses, which we set as correlated at certain time intervals. In this case, these are two synapses on the second dendrite and four synapses on the third dendrite (so there are different numbers of correlated and uncorrelated inputs on different dendrites). After that, when the correlated signal is turned on at the specified time intervals, the selected synapses, with a 90 percent probability for each of them, receive an input identical to the 'mask' input. In this experiment, we use correlated inputs, the correlation coefficient between which is very high (here it is 0.9). Thus, all synapses of the first dendrite, four synapses of the second dendrite, two synapses of the third dendrite, as in the previous experiment, receive low-frequency inputs ($\theta_{input}=0.2$), but two synapses of the second dendrite and four synapses of the third dendrite receive the following inputs: from 200 to 500 steps and from 1400 to 1700 steps are high-frequency inputs ($\theta_{input}=0.5$), and from 800 to 1100 steps and from 2000 to 2300 steps are inputs with $\theta_{input}=0.2$. We include correlated signals at the same time intervals. For the HSS and PPD models, the $\theta_{target}$ is 0.2, and for basic STDP, the $\theta_{target}$ is not assigned. (Figure~\ref{fig:9})

\begin{figure*}
\centering
\includegraphics[width=1.0\textwidth]{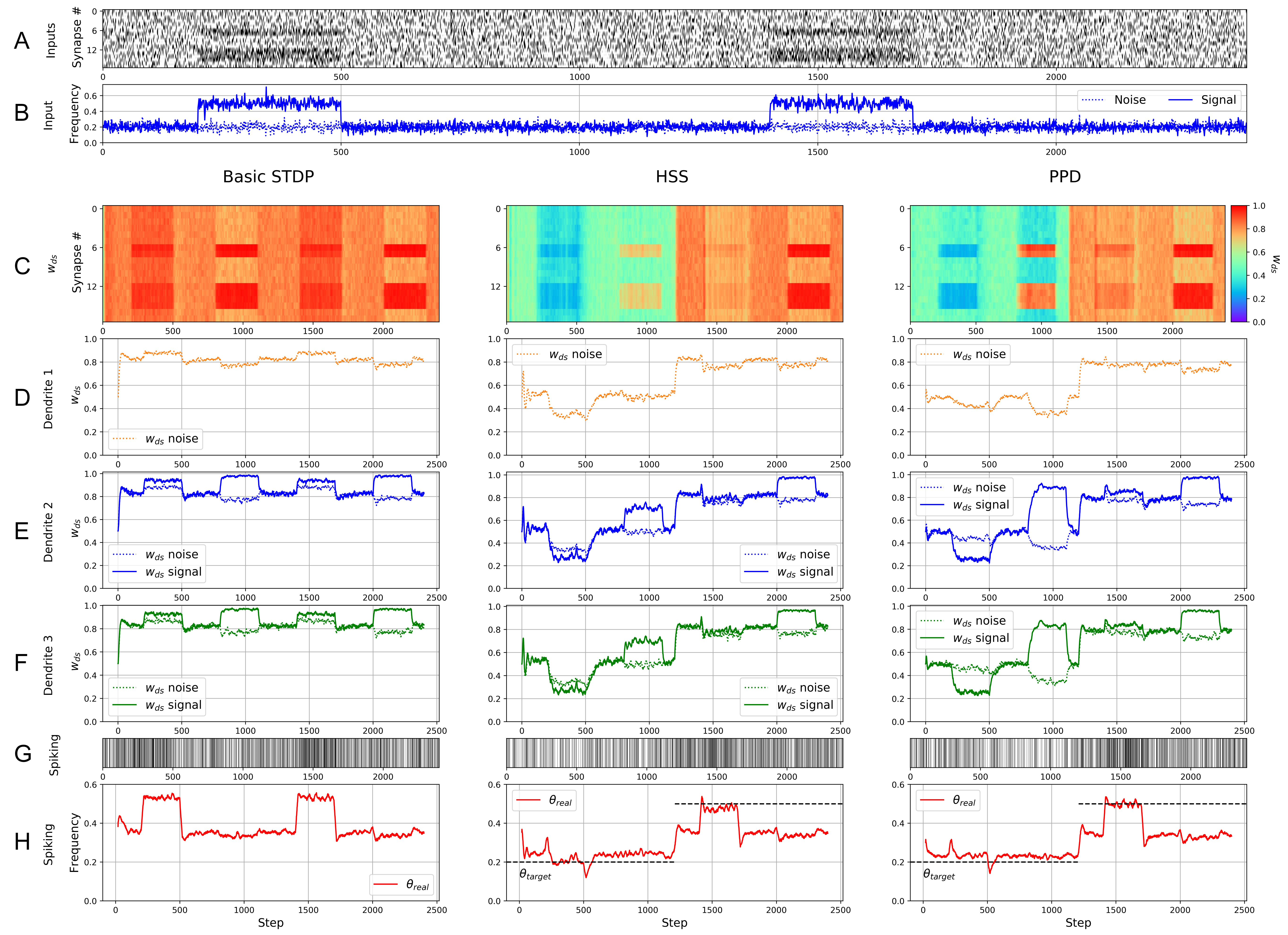}
\caption{Correlated inputs experiments with changing target frequency $\theta_{target}$. All parameters in graphs in the figure are calculated for 2400 steps each of the basic STDP, HSS, and PPD models, respectively, and averaged for all synapses and dendrites over 100 tests for each model (except (A), (G), which are the non-averaged examples from the last test). (A) Inputs with frequency $\theta_{input}=0.2$ arrive at all synapses of the 1st dendrite, 4 synapses of the 2nd dendrite, 2 synapses of the 3rd dendrite. The rest of the synapses (2 synapses of the 2nd dendrite and 4 synapses of the 3rd dendrite) receive the following inputs: from 200 to 500 steps and from 1400 to 1700 steps are high-frequency inputs ($\theta_{input}=0.5$) and correlated signals are included with $\theta_{input}=0.5$, and from 800 to 1100 steps and from 2000 to 2300 steps correlated signals are included with $\theta_{input}=0.2$. Target frequency from 1 to 1200 steps is 0.2, and from 1200 to 2400 steps it is 0.5. (B) Input frequency, where only noise is fed ('Noise') or contains correlated signals ('Signal'). (C) Dynamics of the synaptic weights $w_{ds}$ for all synapses and dendrites,  where the color indicates the $w_{ds}$ value. (D)-(F) Dynamics of the synaptic weights $w_{ds}$ on the 1st, 2nd and 3rd dendrites respectively, '$w_{ds}$ noise' and '$w_{ds}$ signal' are synaptic weights to which only noise is fed ('$w_{ds}$ noise') or contains correlated signals ('$w_{ds}$ signal'). (G) Action potential generation (spiking) $O$ of the neuron ('0' and '1' values) during the work of the models; (H) Spiking frequency $\theta_{real}$ and level of target frequency $\theta_{target}$.}
\label{fig:10}
\end{figure*}

Here we study and compare three models: the basic STDP, the HSS, and the constrained plasticity potential reserve calculation controlled by direct demand (PPD). As can be seen, by heatmaps of all weights (Figure~\ref{fig:9}C), all three models react on frequent stimulus and correlated stimulus in opposite directions. Unlike frequent input, correlated input leads to strong LTP. The basic STDP model has the fastest reaction to signal time and weights' stability for all the models. Nevertheless, it has no firing rate homeostasis. At the same time, both HSS and PPD models do a good job at firing rate maintenance (Figure~\ref{fig:9}H). However, the HSS model does not provide any synaptic competition and heterosynaptic scaling phenomena. When the weights stimulated by correlated signal ('$w_{ds}$ signal') grow, weights with constant input ('$w_{ds}$ noise') are not affect (Figure~\ref{fig:9}D,E,F). In the case of PPD, we may see that LTD accompanies the LTP of correlated synapses in other synapses. At the same time, the LTD of highly stimulated synapses does not affect other synapses' efficiency. It is different from HSS, where LTD occurring in one synapse drives down others. It is important for the network activity that removing one memory does not affect another, but the introduction of new memories leads to higher specificity. In general, PPD has a much better signal-to-noise ratio for such a task. The divergence of differently stimulated versus stable synaptic weights during stimulation calculated as MAE was 0.46 for correlated periods and 0.23 for high-frequency stimulated ones (Table 2). The same values for HSS are 0.13 and 0.19, respectively. We may conclude about almost twice better signal specificity and more useful character of PPD weight dynamics based on this experiment. 

Algorithms of keeping weight homeostasis, such as PPD and HSS, might be helpful also for dynamic network frequency updates and local configuration. In addition, firing-rate regulation might be used for dynamic coupling and efficient information transfer. Thus, it should be examined how homeostatic frequency control approaches to weight restriction would behave with dynamic firing frequency $\theta_{target}$ and how much noise they will transfer with higher $\theta_{target}$. 

We administered some synapses at selected dendrites with correlated inputs of different intensities (high-frequency correlated signal and regular frequency). The synapse order repeats the previous experiment. We also changed the target frequency $\theta_{target}$ from low (0.2) to high (0.5) on two halves of experiment time. 

The basic STDP model has increased all weights of correlated signals. At the same time, both HSS and PPD weight control algorithms filter even strong-correlated signal if it has a high frequency. However, the HSS disadvantage is low specificity because it penalizes all synapses. PPD approach instead penalizes only frequent signals and has more substantial divergence. 

At the high level of target frequency $\theta_{target}$, both HSS and PPD start to react to both input frequencies (Figure~\ref{fig:10}D-F). Nevertheless, regular correlated signals have stronger LTP than high-frequency ones. The experiment with target varying shows the prevalence of the PPD approach over HSS in signal-to-noise ratios. We can see by heatmaps of weights (Figure~\ref{fig:10}C) that weight increase for a correlated signal during the high target period is more distinguishable than during the low target period.  The MAEs for high-frequency and background inputs is 0.06, and for correlated and background is 0.23 for PPD given high the target. The same values for HSS are 0.03 and 0.19 (see Table 2). Both HSS and PPD started to react on noised input when set with high target frequency $\theta_{target}$. We can conclude that both algorithms are suitable for dynamic firing frequency control, but PPD better discriminates between signals.

\begin{table}\label{tab:2}
\caption{Values of the mean absolute error between '$w_{ds}$ noise' and '$w_{ds}$ signal'. STDP, HSS, PPD -- compared models, LT -- low target frequency, HT -- high target frequency, LF -- low input frequency, HF -- high input frequency, FC -- frequent/correlated task, DC -- differently correlated task.}
\begin{tabular}{cclclclc}
\hline
\multicolumn{2}{l}{\multirow{2}{*}{}} & \multicolumn{2}{c}{STDP} & \multicolumn{2}{c}{HSS} & \multicolumn{2}{c}{PPD} \\ \cline{3-8}
\multicolumn{2}{l}{} & \multicolumn{1}{c}{LT} & HT & \multicolumn{1}{c}{LT} & HT & \multicolumn{1}{c}{LT} & HT \\ \hline
\multirow{2}{*}{FC} & LF & 0.20 & \multicolumn{1}{l}{0.19} & 0.20 & \multicolumn{1}{l}{0.19} & 0.46 & \multicolumn{1}{l}{0.23} \\
& HF & 0.06 & \multicolumn{1}{l}{0.07} & 0.05 & \multicolumn{1}{l}{0.03} & 0.18 & \multicolumn{1}{l}{0.06} \\
\multirow{2}{*}{DC} & LF & 0.19 & - & 0.19 & - & 0.46 & - \\
& HF & 0.15 & - & 0.13 & - & 0.23 & - \\ \hline
\end{tabular}
\end{table}

\section{Discussion}\label{sec:5}

The paper proposed the neuronal plasticity regulation model, which changes the potential synaptic weight growth by dynamically controlling the plasticity potential reserve to stabilize the artificial neuron activity. First, we described all parts of the proposed approach, such as the basic plasticity rule, spiking behavior, synaptic plasticity potential reserve dynamics, and the overall paradigm of its regulation. Then, we introduced the new control approaches for regulating plasticity potential storage in the neuron and its dendrites. Finally, we provided four kinds of restricting weight outgrowth methods based on keeping the firing frequency goal. These approaches were tested for reaction on different synaptic input simulations resembling some regimes that might emerge in spiking neural networks. These models also were compared to the performance of the basic STDP weight update rule.

It was shown that a neuron could selectively respond to certain frequencies of input signals depending on the mode of plasticity potential production. A significant feature of the model is that neuron with low-frequency target filters the frequent coincident input while reacting to correlated signals. Neurons of this model are suitable for creating spiking neural networks and can be used for unsupervised training tasks even on modern neuromorphic hardware. The implementations of STDP on memristors \cite{Yao} allow limiting the weight growth in STDP by limiting the voltage-current application time. It resembles the dendritic plasticity potential reserve presented in the article.

Algorithms for homeostatic control of weights that take into account pre-and postsynaptic frequency, proposed in \cite{Liu}, may be interesting for comparison with the model proposed in this article. Such approaches also lead to nonlinear dynamics of weight restrictions, highlighting them from other homeostatic scaling techniques. However, there are no reports on the heterosynaptic synaptic scaling in these papers. \cite{38} implemented heterosynaptic plasticity. Nevertheless, from the original paper, it seems that such a scaling prevents synaptic competition. The approach presented in the current paper, particularly PPD, leads to heterosynaptic scaling, weights, and frequency-dependent homeostasis simultaneously. It is computationally effective and might be implemented not only for one neuron but for the entire layer, thus, reducing the overhead for computing optimal parameters.

Another essential feature of the paper is its high signal-to-noise discrimination ability. Thus, it is beneficial for many tasks for practical use in neural networks. For example, spiking neurons with PPD restricted STDP might be very useful for machine learning tasks.

Future research will be necessary to train the networks of neurons with such rules and introduce some unsupervised learning methods. It might also be useful to find a way to regulate the firing frequency of neurons in the network to minimize the mean firing frequency in the network and maximize the informational transfer. The free-energy principle \cite{Friston} might be helpful for this goal. Also, mutual information measures might be applied to tune SNN with PPD, which will resemble natural brain structures \cite{Schweighofer}. The casualty measures may help obtain the optimal target frequencies for maximal important input-output network transfer \cite{Seth}. In the future, we will also consider turning to a different spiking model because Izhikevich's is limited in firing frequency, and it may be hard to obtain a correct spiking intercity for a given task.

\section*{Declaration of competing interest} The authors declare that they have no known competing financial interests or personal relationships that could have appeared to influence the work reported in this paper.

\section*{Acknowledgement}
The computing resources of the Shared Facility Center "Data Center of FEB RAS" (Khabarovsk) were used to carry out calculations. 

We would like to thank the reviewers and appreciate a lot the valuable comments and corrections given.

\appendix
\section{Appendix}\label{sec:a}

\subsection{Izhikevich model}\label{subsec:a1}

We use the Izhikevich model \cite{Izhikevich} as the activation function for calculation of spike output $O$ of the neuron, which is formulated as follows:

\begin{equation}\label{eq:a1}
\begin{array}{ll}
\dot v = 0.04v^2+5v+140-u+I, \\
\dot u = a(bv-u)
\end{array}
\end{equation} 

In the case of membrane voltage $v$ exceeding the threshold $v_{peak}$, the following resetting happen:

\begin{equation}\label{eq:a2}
\text{if } v \geq v_{peak} \text{, then }
\begin{cases}
v \leftarrow c \\
u \leftarrow u+d 
\end{cases}
\end{equation}

Here, $u$ is the membrane recovery variable, providing negative feedback to $v$; $I$ is the input current weighted on synapses and integrated over dendrites; $a$ is the time scale of the recovery variable $u$; $b$ is the sensitivity of the recovery variable $u$ to the dynamic of membrane potential $v$; $c$ is the $t$ value to which the membrane potential $v$ is resettled in case of the spike, and $d$ is the after-spike reset value of $u$.

As we described in Section~\ref{sec:2}, we compute the input current $I$ by averaging across synapses and dendrites. The average for synapses is considered to ensure that the number of synapses and dendrites does not affect the dynamics of the activation function (Equation~\ref{eq:1}). We scale the input activity by multiplying on scaling coefficient $k^{izh}$. 

To find the coefficient $k^{izh}$, consider a particular case when $\theta_{input}$ is 0.5, so the average input current $I_{ds}$ is also 0.5, and the weights $w_{ds}$ are in its middle state ($\frac{w_{max} - w_{min}}{2}$) and should lead to $\theta_{real}$ equal to 0.5. Thus, we can substitute these values to Equation~\ref{eq:1} ($w_{max} = 1$, $w_{min} = 0$):

\begin{equation}\label{eq:a3}
I = \bar{I} \cdot k^{izh} = \frac{2\cdot \theta_{input} \cdot \frac{(w_{max} - w_{min})}{2}}{(w_{max} - w_{min})}\cdot k^{izh} = \theta_{input} \cdot k^{izh}.
\end{equation}

The number of dendrites $D$ and synapses $S$ in this formula can be neglected, since we take averaged input and weights. So, $\bar{I}=\theta_{input}=0.5$, and the first line in Equation~\ref{eq:a1} will be the following:

\begin{equation}\label{eq:a4}
\dot v = 0.04v^2 + 5v + 140 - \hat{u} + \bar{I} \cdot k^{izh}.
\end{equation}

Here, $\hat{u}$ is the highest possible value of the $u$ when $I=\bar{I}$ and $\theta_{real}=0.5$.

\begin{equation}\label{eq:a5}
\bar{I} \cdot k^{izh} = \dot v - 0.04v^2 - 5v - 140 + \hat{u}.
\end{equation}

Since we want spikes to occur every $\tau$ steps, this means that $\dot v$ must change from $c$ to $v_{peak}$ per $\tau$ steps. Then the values of $v$ and $\dot v$ will take the form $v = c$, $\dot v = \frac{v_{peak} - c}{\tau}$, and $k^{izh}$ will be

\begin{equation}\label{eq:a6}
k^{izh} = \frac{\frac{v_{peak} - c}{\tau} - 0.04c^2 - 5c - 140 + \hat{u}}{\bar{I}}.
\end{equation}

Further, to find the coefficient $k^{izh}$, we need to get the highest possible value of the $u$, as it is always a downward value for the $\dot v$. According to Equation~\ref{eq:a2}, each time the spike occurs, $d$ is added to the value of the $u$. Thus, we add $d$ every $\tau$ steps, where $\tau = \frac{1}{\theta_{real}} = 2$:

\begin{equation}\label{eq:a7}
u(t) = u(t-1) + a(bv - \hat{u}) + \frac{d}{\tau}.
\end{equation}

Since we are looking for the maximum value of $u$, and by the nature of the function, it is such when $u(t) - u(t-1) = 0$, Equation~\ref{eq:a7} takes the form

\begin{equation}\label{eq:a8}
0 = abv - a\hat{u} + \frac{d}{\tau},
\end{equation}

\begin{equation}\label{eq:a9}
\hat{u} = bv + \frac{d}{a \tau},
\end{equation}

We substitute the value of $v$ in Equation~\ref{eq:a9} with

\begin{equation}\label{eq:a10}
\hat{v} = c + \frac{v_{peak} - c}{\tau},
\end{equation}
where $\hat{v}$ is the average value of $v$ at which $u$ becomes the maximum.

Thus, the relation of $\hat{u}$ takes the form

\begin{equation}\label{eq:a11}
\hat{u} = b(c + \frac{v_{peak} - c}{\tau}) + \frac{d}{a \tau},
\end{equation}

and $\bar{I}=0.5$, $\tau = 2$, so with simple transformations we can define the coefficient $k^{izh}$ from Equation~\ref{eq:a6} as

\begin{equation}\label{eq:a12}
k^{izh} = (1+b)\cdot v_{peak}+\frac{d}{a}-0.08c^2+(b-11)\cdot c-280,
\end{equation}
where $a$, $b$, $c$, $d$, $v_{peak}$ are the same parameters that are presented in Equations~\ref{eq:a1}--\ref{eq:a2}.

In our model the parameters of the spiking neuron model are as follows: $a=0.02$, $b=0.23$, $c=-65.0$, $d=2.00$, $v_{peak}=20 \text{ mV}$.

This scaling coefficient is used to adapt the input to the Izhikevich model to allow the firing frequency variation necessary for our experiments.

\subsection{Spike-timing-dependent plasticity (STDP)}\label{subsec:a2}

The STDP model describes the change in synaptic weight as a function of the delay between the timing of the release of spikes by presynaptic and postsynaptic neurons. We denote $\Delta w^{stdp}_{ds}$ as an update of the synaptic weight of the synapse $s$ on the dendrite $d$, according to the model, introduced in \cite{STDP}:

\begin{equation}\label{eq:a13}
\Delta w^{stdp'}_{ds} =
\begin{cases}
A_+(w_{ds})\cdot \exp(-\Delta t/\tau_+), & \text{if }\Delta t\ge0,  \\
-A_-(w_{ds})\cdot \exp(\Delta t/\tau_-), & \text{if }\Delta t<0,
\end{cases}
\end{equation}
where $\Delta w^{stdp'}_{ds}$ is the update of the synaptic weights by STDP rule; $\Delta t = t_{sp}^{(pre)} - t_{sp}^{(post)}$ is the interval between the occurrence of subsequent spikes at presynaptic and postsynaptic neurons; $\tau_+$ and $\tau_-$ is the characteristic time defining a time interval, which may cause successful correlation effects; $A_+(w_{ds})$ and $A_-(w_{ds})$ are the coefficients that reflect the maximum possible rate of change of synaptic weights $w_{ds}$ and may depend upon its current value.

To ensure that the synaptic weights do not take abnormal values during the dynamics of the neural network, the interval $w_{min} < w_{ds} < w_{max}$ of their admissible values is usually introduced and, respectively, the functions $A_+(w_{ds})$ and $A_-(w_{ds})$ are defined as follows:

\begin{equation}\label{eq:a14}
\begin{array}{lcl}
A_+(w_{ds}) = (w_{max} - w_{ds})\cdot A_+, \\
A_-(w_{ds}) = (w_{ds} - w_{min})\cdot A_-,
\end{array}
\end{equation}
where $A_+$ and $A_-$ are the positive parameters.

Here, in our model: $A_+=0.25$, $A_-=0.25$, $w_{max}=1.0$, $w_{min}=0.0$, $\tau_+=10$, $\tau_-=10$.  

\bibliographystyle{unsrt}

\bibliography{cas-refs_revision}

\end{document}